\documentstyle[12pt,epsf]{article}
\begin{document}

\title{Quasiperiodic functions and Dynamical Systems
in Quantum Solid State Physics.}

\author{A.Ya.Maltsev$^{1}$, S.P.Novikov$^{1,2}$.
\footnote{The work of S.P.Novikov is partially supported
by the NSF Grant DMS 0072700.}}
 
\date{
\centerline{$^{(1)}$ L.D.Landau Institute for Theoretical Physics,}
\centerline{119334 ul. Kosygina 2, Moscow, }
\centerline{ maltsev@itp.ac.ru \,\, ,
\,\,\, novikov@itp.ac.ru}
\centerline{$^{(2)}$ IPST, University of Maryland,}
\centerline{College Park MD 20742-2431,USA}   
\centerline{novikov@ipst.umd.edu}}

\maketitle

\begin{abstract}
 This is a survey article dedicated to the study of topological 
quantities in theory of normal metals discovered in the works of 
the authors during the last years. Our results are based 
on the theory of dynamical systems on Fermi surfaces.
The physical foundations of this theory (the so-called "Geometric
Strong Magnetic Field Limit") were found by the school of 
I.M.Lifshitz many years ago. Here the new aspects in the 
topology of quasiperiodic functions are developed.
\end{abstract}

\centerline{\bf Introduction. Quasiperiodic functions and 
Dynamical Systems.}

\vspace{0.2cm} According to the standard definition, quasiperiodic
function of $l$ variables with $m$ quasiperiods is  a restriction
of any periodic function $f(y_1,\ldots,y_m)$ of $m$ variables on
the $l$-dimensional affine subspace $R^l    \subset R^m$.

{\bf Problem}. What can one say about  topology of the level
curves of the quasiperiodic functions on the plane $R^2$?

As we are going to demonstrate below, this problem has very
important physical interpretation for the case where the number of
quasiperiods is equal to m=3. These studies motivated by the solid
state physics, were started by S.P.Novikov in 1982 (see \cite{nov1}).
They were  continued in  his seminar 
(see \cite{zorich,nov2,nov3,dynn3}) as a
nice purely topological investigations. In particular, an
important breakthrough in the purely topological aspects of the
"Novikov problem" for $m=3$ (i.e. exactly for the dimensions
needed in physics) has been made by A.V.Zorich (see\cite{zorich}) 
and I.A.Dynnikov (\cite{dynn3}). 
Long period no physical applications were
expected. However, it was found later by the present authors (see
\cite{novmal1,novmal2}) that topological characteristics of this 
picture discovered in the previous works are organized in some 
sort of "topological resonance" leading to the important physical
conclusions for the electric conductivity in the strong magnetic
field. We shall explain these results below.

 Let us explain a deep connection of the theory of quasiperiodic
functions with special hamiltonian systems. Consider a torus $T^m$
as a factor-space of the Euclidean  space $R^m$ with coordinates
$p_1,\ldots,p _m$ by the lattice $\Gamma^*$ of the rank $m$. Let a
constant Poisson Bracket be given on 
the torus by the skew-symmetric matrix
$B^{ij}=-B^{ji}$ with rank equal to $2s$. There are exactly
$q=m-2s$ linear "Casimirs" or "annihilator" functions
$l_1,\ldots,l_q$ such that

$$\{l_a,f(p)\}=B^{ab}\partial_i l_a \partial_b f=0$$

 Any hamiltonian function $\epsilon({\bf p})$ 
on the torus $T^m$ determines a flow
$$dp_i/dt=\{p_i,\epsilon\}=B^{ab}\partial_ap_i\partial_b\epsilon$$
This system has only one well-defined  one-valued integral of
motion $\epsilon=const$ and $q$ multivalued conservative
quantities $l_j=const,j=1,\ldots,q$ generating a foliation on the
torus. Let us point out that the restriction of the function
$\epsilon({\bf p})$ on the leaf of this foliation is exactly a
quasiperiodic function with $m-q=2s$ variables and $m$
quasiperiods.

 We concentrated our studies on the special case  $s=1$, i.e. rank
of the Poisson Bracket is minimal possible (equal to 2). Here we
have $2s=2$, i.e. our quasiperiodic functions are defined on the
2-planes belonging to the family of  parallel 2-planes in the
space
 $R^2_l\subset R^m,l=l_1\ldots,l_{m-2}$ where $l_j=const,j=1,\ldots,m-2$.
We call energy levels by the ''Fermi Surfaces'' in this case. The
restriction of the straight-line foliation on the surface
$\epsilon=const$ gives trajectories of our hamiltonian system.
They are obtained as sections of  Fermi Surface by the planes
$l_j=const,j=1,\ldots,m-2$. We arrived to the following

{\bf Conclusion.} Trajectories of the hamiltonian system described
above exactly coincide with  projections of the level curves of
the quasiperiodic functions on the planes $R^2_l\subset R^m$ under
the covering map $R^m\rightarrow T^m $.

As we shall see in the paragraph 2, the case $m=3,q=1$ plays
fundamental role in the solid state physics. The case $m=4,s=1$
also has been investigated. Some partial results were obtained in
the work \cite{nov5}.
 \newtheorem{de}{Definition}
\begin{de}
We call trajectory compact if it defines a compact curve on the
plane $R^2$ as a level curve of the quasiperiodic function.
\end{de}

It follows from definition that periodic trajectories on  Fermi
surfaces in the 3-torus are compact if and only if they are
homotopic to zero in the torus.

\begin{de}
We call trajectory quasiperiodic with average direction $\eta$ if
and only if corresponding level of quasiperiodic function
$\epsilon=const$ on the plane $R^2_l$ lies in the strip of finite
width between two  straight lines parallel to the direction
$\eta$. All trajectories more complicated than compact and
quasiperiodic, we call chaotic.
\end{de}
In particular, all periodic trajectories non-homotopic to zero in
the torus $T^m$, are quasiperiodic in that definition.

\begin{de}
We call hamiltonian system "topologically completely integrable"
on the given energy level if and only if all its trajectories are
either compact or quasiperiodic on this level $\epsilon=const$.
\end{de}
This definition has nothing common with the standard Liouville
Complete Integrability.
\begin{de}
We call quasiperiodic function topologically completely integrable
if and only if all its levels are either compact or quasiperiodic.
We call it stably completely integrable if it remains
topologically completely integrable after any $C^{\infty}$-small
perturbation of the m-periodic function $\epsilon(p_1,\ldots,p_m)$
and any small perturbation of the linear functions
$l_j,j=1,\ldots,m-2$.
\end{de}

 This set of definitions gives only the first impression for the
topological properties really needed for the physical
applications. As it was found in the works 
\cite{novmal1,novmal2} where
physical applications were obtained, there is a remarkable
''topological resonance'' of the topological quantities here
extracted from the heart of the proofs of the Zorich and Dynnikov
theorems, leading to the applications including the new observable
topological phenomena:

{\bf Topological Resonance}
was found in \cite{novmal1} for the applications in physics. 
Quasiperiodic functions should be
stably completely integrable for all family of parallel directions
$l$. All quasiperiodic trajectories in this family should have the
same average direction $\eta$. This direction should define an
integral 2-plane $\eta\in Z^2$ in the reciprocal lattice
$Z^3=\Gamma^*$. This integral 2-plane should be rigid under the
small perturbations defining an open "Stability zone" on the
sphere $S^2$ for the case $m=3,r=1$ where this plane is the same.

This topological resonance is valid for the generic directions, it
is certainly untrue for the rational directions of magnetic field.
The same definition works also for higher dimensions replacing
2-plane by the $(n-1)$-plane.

  We shall return to the explicit
formulation of corresponding results later. The exposition of
topological theorems in the final form convenient for applications
can be found with full set of proofs in \cite{dynn7} for the case
$m=3,s=1$. Only partial results were obtained for the case
$m=4,s=1$ (see\cite{nov5}). As it was established in these cases,
our hamiltonian system is generically stably completely
integrable. Let $m=3,s=1$. More precisely, for the generic
nonsingular Fermi surface $\epsilon=const$ the set of linear forms
$l_1\in S^2$ ("directions of the magnetic fields" in physics)
with chaotic dynamics has a Hausdorf dimension
  no more than 1 on the 2-sphere. In particular, its measure is equal
to zero as it was proved by Dynnikov in 1999.

{\bf Conjecture.} A Hausdorf dimension of the chaotic cases is
strictly less than 1 for the physical case $m=1,s=1$.

 For $m=4,s=1$ we presented an idea of the proof  in 
\cite{nov5} that the "stably integrable set" is open
and dense in the Grassmanian $(l_1,l_2)\in G_{2,2}$ if the generic
"Fermi level" $\epsilon=const$ is fixed in the space $T^4$.
However, nothing is known about its measure. Our conjecture is
that its measure is full, but no idea of the proof is known now.

We think that for $m\geq 5, s=1$ these systems are generically
chaotic.

\section{Normal Metals: The Standard Model.}.

\vspace{0.2cm}

 As it was well-known  many years,
even the ordinary electrical conductivity in single crystal
normal metals cannot be explained properly without quantum
mechanics. A working model for studying it has been elaborated in
1930s based on the quantum states of the free electron Fermi gas
in some external 3D periodic potential created by the lattice
$\Gamma$ of ions in the Euclidean space $R^3$. For the zero
temperature $T=0$ our system lives in the standard Dirac-type
ground state where all states with lowest energies are occupied
(one electron for one state). The states of electron correspond
exactly to the Bloch eigenfunctions of the one-particle
Schr\"odinger operator $L\psi=\epsilon\psi$. By definition, 
we have

$$\psi({\bf x} + {\bf a}) = 
\exp\{i<{\bf a},{\bf p}>/\hbar\} \, \psi(x) \,\,\, ,
\,\,\,\,\, {\bf a} \in \Gamma$$

 They depend on the "quasimomentum vector" ${\bf p}$ 
(or wave vector ${\bf k} = {\bf p}/\hbar$) belonging 
to the  "Space of Quasimomenta" 

$${\bf p} \in T^{3} =R^{3}/\Gamma^{*}$$ 
where $\Gamma^{*}$ is a reciprocal lattice dual to the 
lattice $\Gamma$, i.e. for ${\bf p} \in \Gamma^{*}$ we have
 $$<{\bf a},{\bf p}>=2\pi \hbar n \,\,\, , \,\,\,\,\, n\in Z$$
 For the given value of  quasimomenta $p\in T^3$ there is a
discrete spectrum of real energies 
$\epsilon_n({\bf p}), n\geq 0$. They
are called "Dispersion relations". According to the Fermi
statistics, all levels $\epsilon < \epsilon_F$ are occupied for the
zero temperature where the exact value  $\epsilon_F$ (the "Fermi
Energy") depends on the number of free electrons in the metal. 
 We assume that this
"Fermi level" $\epsilon({\bf p})=\epsilon_F$ is a nondegenerate
2-manifold $S_{F}$ in the 3-torus of quasimomenta. It is called 
{\bf Fermi Surface}. Physicists normally draw all pictures in the
universal covering space $R^3 \rightarrow T^3$ where we have a
covering Fermi surface $\hat{S}\rightarrow S_{F}$ presented as a
level of periodic function $\epsilon({\bf p})$ of 3 variables
$p_1,p_2,p_3$. We call it "a periodic surface". A fundamental
("Dirichle") domain of the lattice $\Gamma^*$ in $R^3$ people
call "the first Brillouen zone" in the physics literature. For the
temperature low enough the excited electron states are located in
the small area nearby of Fermi level, so we continue to use a
geometric picture described above. The limits for temperature
should be discussed later. We are arriving to the conclusion that
{\bf in the standard low-temperature model of normal metal the set
of active  electrons are identified with points of Fermi Surface
$S_{F}$ in the Space of Quasimomenta $T^3$.} This surface is
orientable and homologous to zero in the 3-torus. Its topology can
be complicated in some important cases like copper, gold, platinum
and others. The most important topological characteristic of the
connected piece of  Fermi Surface is the {\bf Topological rank}
(introduced in \cite{novmal2}).

\begin{de}
 By the "Topological rank" $r$ of the connected  piece  of  Fermi
 Surface we call a
rank of the image-lattice $\pi_1(S_{F})\rightarrow \pi_1(T^3)=Z^3$.
Obviously we have $r=0,1,2,3$. We call connected Fermi surface
topologically complicated if its topological rank is equal to $3$.

\end{de}

 Let us mention here that the first experimental observation
of the Fermi surface corresponding to the Topological Rank $3$
were made by Pippard for $Cu$ (\cite{pippard}).

\newtheorem{lem}{Lemma}
\begin{lem}
For any connected piece of Fermi Surface homologous to zero
following inequality is true $$g\geq r$$
 where $r$ and $g$ are the
topological rank and genus of Fermi surface correspondingly. For
any connected piece of Fermi Surface non-homologous to zero the
topological rank is no more than 2.

\end{lem}

Proof. For $r=0,1$ the first case of this lemma is trivial. To
prove it for $r=2,3$ we need to use that the homology class
$i_*[S_{F}]\in H_2(T^3)$ is equal to zero for the embedding
$i:S_{F}\rightarrow T^3$. Therefore the cohomological homomorphism
$H^2(T^3)\rightarrow H^2(S_{F})$ is also trivial. Therefore the
product of any pair of classes $$y,z\in i^*(H^1(T^3))=Z^r\subset
H^1(S_{F})=Z^{2g}$$ is equal to zero. The space $H^1(S_{F})=Z^{2g}$ 
is symplectic nondegenerate, and its subspace $i^*(H^1(T^3)=Z^r$ is
Lagrangian. Therefore we have $$2g\geq 2r$$

 Consider now any piece non-homologous to zero in the 3-torus. 
This submanifold $N_F^0$ in the 3-torus is orientable. There exist
1-dimensional cycle in the 3-torus having nonzero intersection
index with it. This cycle and its multiples do not belong to the
image of the map $i_*:H_1(S_{F}^0)\rightarrow H_1(T^3)=Z^3$.
Therefore its rank is no more than 2.

Our lemma is proved.

Let us remind here that for such noble metal as gold we have $r=3,
g=4$.

\section{Electrons in the magnetic field. Dynamical
Systems on Fermi Surfaces}

 The classification of states described above works well in the
absence of magnetic field. It is very difficult to study
Schroedinger equation in the presence of magnetic field combined
with   periodic lattice potential. Nobody succeeded to find any
suitable classification of the one-electron states if magnetic
flux through elementary cell is irrational (in the natural quantum
units) even for 2D crystals. For the real natural 3D crystals the
size of elementary lattice is about $10^{-16}cm^2$. Therefore even
the strong magnetic field of the order $1t\sim 10^4 Gauss$ gives
only a small fraction of the quantum unit (about $10^{-3}$). Many
years ago physicists developed a''semiclassical approach'' to this
problem where the dispersion relations and zone structure is taken
exactly from the quantum theory and magnetic field is added
"classically". The leading role in these studies since 1950s
played Kharkov group of I.M.Lifshitz and his pupils (M.Azbel,
M.Kaganov, V.Peschanski and others 
\cite{lifazkag}-\cite{lifpes2}, 
see also \cite{lifkag1}-\cite{abr}). It simply
means that electrons start to move along the Fermi Surface in the
space of quasimomenta ${\bf p} \in T^3 = R^3/\Gamma^{*}$ 
with Poisson Bracket determined by the magnetic field

$$\{p_i,p_j\} \, = \, {e \over c} \, B^{ij}$$
where $\epsilon_{ijk} B^{ij} = B_{k}$ is the
ordinary vector of magnetic field dual to the skew symmetric
tensor $B^{ij}$. This motion is generated by the hamiltonian
$\epsilon({\bf p})$ equal to the dispersion relation. We have

\begin{equation}
\label{dynsyst}
dp_i/dt \, = \, {e \over c} \, B^{ij} \, 
\partial_j\epsilon({\bf p}) \, = \,
{e \over c} [\nabla \epsilon({\bf p}) \times {\bf B}]
\end{equation}

 This system can be easily integrated analytically: in particular,
its trajectories are exactly sections of the Fermi surface $S_{F}$
given by  equation $\epsilon = \epsilon_F$, by the plane orthogonal
to the vector ${\bf B} = (B_1,B_2,B_3)$ of  magnetic field.

{\bf However, one should not think that this system is trivial
because we have to identify points of the Euclidean space
equivalent modulo reciprocal lattice}. There are examples
(constructed by S.Tsarev and I.Dynnikov--see in the survey article
\cite{dynn4})  such  that this system is chaotic.

 Physicists of the Lifshitz group mentioned above formulated (and
verified on the physical level) following fundamental

{\bf Geometric Strong Magnetic Field Limit}: All essential
properties of the electrical conductivity in the presence of the
reasonably "strong" magnetic field (however, not exceeding the
limits of semiclassical approximation) depend in main approximation 
on that dynamical system only.

The exposition of Kinetic Theory arguments leading to this
conclusion can be found in the next paragraphs. In particular,
these arguments lead to the values of magnetic field between one
and several hundreds $t$ ($1t=10^4Gauss$) for the real crystals
like gold and temperatures like $T\sim 1K$.

\section{Electron dynamics and Topological phenomena.}

 Let us consider now in more details
the physical phenomena connected with the 
geometry of quasi-classical electron orbits on the Fermi-surface.
Namely, we are going to deal with the conductivity in normal
metals in the presence of the strong homogeneous magnetic field
${\bf B}$. Let us explain first the concept of geometrical
limit in this situation.

 According to the standard approach we use the one-particle
distribution function $f_{s}({\bf p})$ defined on the 
three-dimensional torus $T^{3}$ for every energy band $s$.
The values of the functions $f_{s}({\bf p})$ always belong
to the interval $[0,1]$ for the fermions and the number of
particles occupying the volume element $d^{3}p$ in the energy
band $s$ can be written as

$$d^{3} N_{s} = 2 \, f_{s}({\bf p}) \, 
{d^{3} p \over (2\pi\hbar)^{3}} \, V$$
where $V$ is the total volume and the multiplier $2$ is
responsible for the spin degeneration. For the concentration
of particles $n = N/V$ the analogous formula can then be
written as

$$d^{3} n_{s} = 2 \, f_{s}({\bf p}) \,
{d^{3} p \over (2\pi\hbar)^{3}} \, $$

 In the absence of the external fields any distribution
function $f_{s}({\bf p})$ can be written in the form of
well-known Fermi distribution corresponding to some fixed
temperature $T$:

\begin{equation}
\label{fermdist}
f^{T}_{s}({\bf p}) = 
{1 \over 1 + exp({\epsilon_{s}({\bf p}) - \epsilon_{F} \over T})}
\end{equation}

 The parameter $\epsilon_{F}$ is called the Fermi energy of  
metal and can be defined from the total concentration of
particles and the form of dispersion relations 
$\epsilon_{s}({\bf p})$. 

 We are going to consider the situation of rather small 
temperatures $T \sim 1K$ with respect to 
the width of the energy bands
(usually $\epsilon_{max} - \epsilon_{min} \sim 10^{4}-10^{5} K$).
We can put then $f_{s}({\bf p}) \equiv 1$ for the energy bands
lying completely below the Fermi energy $\epsilon_{F}$ and
$f_{s}({\bf p}) \equiv 0$ for the bands lying completely above
$\epsilon_{F}$. Easy to see that this property will be conserved
also in the presence of small external perturbations by energy 
reasons and only the bands with 
$\epsilon_{min} < \epsilon_{F} < \epsilon_{max}$ 
(conductivity bands) can be
interesting for us. As can be easily seen from (\ref{fermdist})
all the functions $f_{s}({\bf p})$ are smooth for $T > 0$
and change rapidly from $0$ to $1$ in the narrow region $\sim T$
near the Fermi level $\epsilon_{s}({\bf p}) = \epsilon_{F}$
(see Fig. 1)

\begin{figure}
  \begin{picture}(0, 170)
    \leavevmode
    \put(0,0){\epsfxsize=1.0\hsize \epsfbox{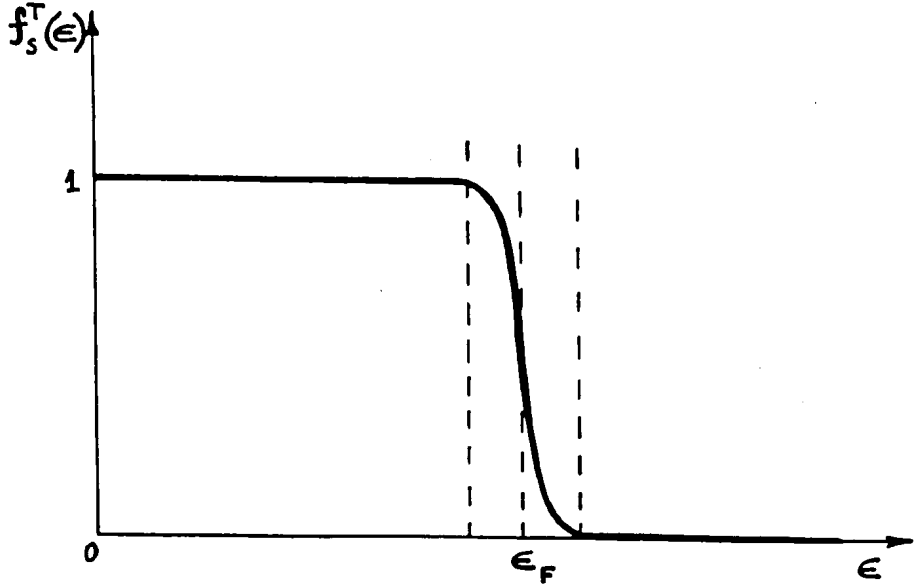}}
  \end{picture}
  \caption{The Fermi distribution.}
\end{figure}

 For $T = 0$ we can put formally

$$f_{s}({\bf p}) = \left\{ \begin{array}{l}
1 \,\,\,\,\, {\rm if} \,\,\,\,\,
\epsilon_{s}({\bf p}) < \epsilon_{F} \cr
0 \,\,\,\,\, {\rm if} \,\,\,\, 
\epsilon_{s}({\bf p}) > \epsilon_{F} 
\end{array} \right. $$
though the zero temperature can not be obtained in the real
experiments.

 The total electric current for any given electron distribution
can be calculated as the integral of group velocity
${\bf v}^{gr}_{s}({\bf p}) = \nabla \epsilon_{s}({\bf p})$
over all the energy bands with the weights given by functions
$f_{s}({\bf p})$. We have so

$${\bf j} = 2 \sum_{s}\mathop{\int\dots\int}\limits_{T^3} 
e {\bf v}^{gr}_{s}({\bf p}) f_{s}({\bf p}) 
{d^{3} p \over (2\pi\hbar)^{3}} =$$

$$= 2 e \sum_{s}\mathop{\int\dots\int}\limits_{T^3}
\nabla \epsilon_{s}({\bf p}) f_{s}({\bf p})
{d^{3} p \over (2\pi\hbar)^{3}} $$

 It is easy to see that all the functions 
${\bf v}^{gr}_{s}({\bf p})$ are the odd functions on the tori
$T^{3}$ $(\epsilon_{s}({\bf p}) = \epsilon_{s}(-{\bf p}))$
and the total electric current is zero for any Fermi distribution
$f^{T}_{s}({\bf p})$ given by (\ref{fermdist}).

 In our quasiclassical approach we will neglect the quantization
of the electron energy levels in the magnetic field ${\bf B}$ and 
use just the classical system (\ref{dynsyst}) to describe the
electron behavior in the presence of magnetic field. Let us just
point out here that for $\hbar \omega_{B} \ll \epsilon_{F}$ the
quantization will not change the geometric characteristics
of conductivity for rather strong magnetic fields.

 Let us note now that the dynamical system (\ref{dynsyst}) does
not change any distribution (\ref{fermdist}) since both the
energy $\epsilon_{s}({\bf p})$ and the volume element $d^{3} p$
are conserved by this system. However, the form of the linear
response to the small electric field ${\bf E}$ depends 
strongly on the geometry of the trajectories of (\ref{dynsyst})
as we will see below.

 We define the Fermi surface $S_{F}$ as the union of all surfaces
given by the equation $\epsilon_{s}({\bf p}) = \epsilon_{F}$ 
for all conductivity bands. It can be shown that these
parts do not intersect each other in the generic situation
by quantum mechanical reasons and the Fermi surface can be 
usually represented as a disjoint union of smooth compact
nonselfintersecting pieces in the three-dimensional torus
$T^{3}$. The Fermi surface $S_{F}$ is homologous to zero in $T^{3}$ 
by construction and all the components
of $S_{F}$ give the independent contribution to the conductivity
tensor $\sigma^{ik}$. We can consider then separately every 
conductivity zone with the dispersion relation 
$\epsilon({\bf p}) = \epsilon_{s}({\bf p})$. The corresponding
contributions to the conductivity should then just be added
in the three dimensional tensor $\sigma^{ik}$.  
The forms of these contributions will however use also
the fact of nonselfintersecting of different components of
the Fermi surface as will follow from the Topological consideration.
We will discuss later also the case when for some special reason
different components of $S_{F}$ can intersect each other.

 Let us describe now the "Geometric Strong Magnetic Field limit" of 
conductivity in our situation. 

 We will introduce the mean electron free motion time $\tau$ 
characterizing the mean time interval of the free electron
motion between the two scattering acts. The time $\tau$ is
defined just by scattering on the impurities for rather low
temperatures $T$ and depends on the purity of the crystal.
We can assume then that every electron lives on the same 
trajectory of (\ref{dynsyst}) during the mean time $\tau$ 
and change the trajectory after the scattering act. The 
geometric length of the corresponding trajectory interval 
$l_{p}$ (in ${\bf p}$-space) is proportional to the magnetic 
field $B$ and tends to the infinity as 
$B\tau \rightarrow \infty$. We can see then that any small
perturbation of the Fermi distribution (\ref{fermdist})
will be instantly "mixed" along the trajectories of 
(\ref{dynsyst}) in this strong magnetic field limit. All
the averaged values should thus then be calculated in this 
situation for the "averaged" distribution constant on the 
trajectories of (\ref{fermdist}). All stationary distributions
should also be constant on the trajectories of (\ref{dynsyst})
as $B\tau \rightarrow \infty$ and can just be slightly different
from these constants for large finite values of $B\tau$.
As was shown in ~\cite{lifazkag} these distributions can be
expanded as regular functions in powers of $(B\tau)^{-1}$
in this limit for closed and open periodic electron orbits.
This approach will work well also for more general cases of
regular stable open orbits giving the similar effects in
this case. However, as we will see below, it can not be applied
in more complicated "chaotic" behavior of
the electron orbits and the situation is much more complicated
in this case.

 We will call the {\bf Geometric Strong Magnetic Field Limit }
the situation when 
$l_{p} \gg p_{0}$ where $p_{0}$ is the size of the Brillouen
zone in the ${\bf p}$-space. For the crystal lattice with the
lattice constant $\sim a$ we will have $p_{0} \sim 2\pi\hbar/a$.
Let us introduce the cyclotron frequency 
$\omega_{B} = eB/m^{*}c$ where $m^{*}$ is the "effective mass"
of electron in the crystal. Using the standard approximation
$m^{*} v_{gr} \sim p_{F} \sim p_{0}$ on the Fermi surface we
can write the condition of strong magnetic field limit in usual
form $\omega_{B} \tau \gg 1$.

 The electron dynamics in ${\bf x}$-space can be described by 
additional system

$${\dot {\bf x}} = {\bf v}_{gr}({\bf p})$$
(in quasiclassical approach) and can be easily reconstructed
for any known trajectory of (\ref{dynsyst}) in ${\bf p}$-space.
Let us choose now the $z$-axis along the direction of
the magnetic field ${\bf B}$ while the $xy$-plane will be
orthogonal to ${\bf B}$. It is easy to see then that the $xy$
projection of the electron trajectory in ${\bf x}$-space 
can be obtained just by rotation of corresponding trajectory
of (\ref{dynsyst}) by $\pi/2$. As a corollary of this fact the
asymptotic behavior of conductivity in the plane orthogonal
to ${\bf B}$ is defined completely by the geometry of the
electron orbits in ${\bf p}$-space 
(\cite{lifazkag}-\cite{lifpes2}). 

 Let us consider here the closed and the open periodic electron
trajectories in the ${\bf p}$-space (\cite{lifazkag}).
As easy to see the closed trajectories can arise in many 
different situations. The open periodic trajectories 
can be obtained
for instance as the intersection of the periodic cylinder
in ${\bf p}$-space by the plane containing the vector parallel
to the axis of cylinder (see Fig 2. a,b).

\begin{figure}
  \begin{picture}(0, 170)
    \leavevmode
    \put(0,0){\epsfxsize=1.0\hsize \epsfbox{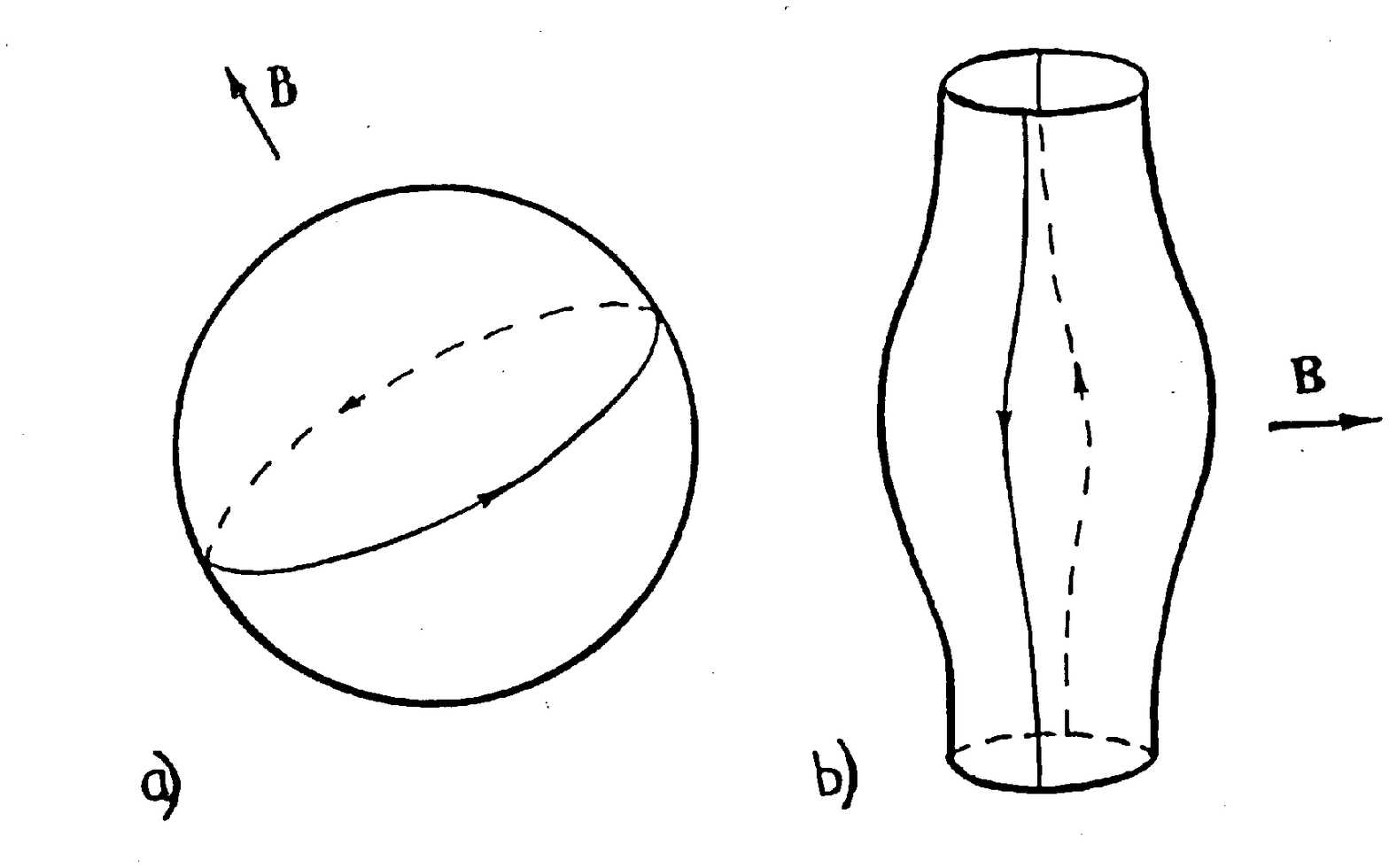}}   
  \end{picture}
  \caption{The simplest closed and open periodic trajectories.}
\end{figure}

 We note here that the open periodic trajectories always come
in pairs with opposite parallel directions as follows from the
fact that the Fermi surface is homologous to zero in $T^{3}$.
 
 Let us choose the $x$-axis along the mean direction of the open
orbits in ${\bf p}$-space in the plane orthogonal to ${\bf B}$ 
for the second
situation and arbitrarily in the plane orthogonal to ${\bf B}$
for the case of closed electron orbits only.

 The projection of the mean direction of open orbits in the
plane orthogonal to ${\bf B}$ in ${\bf x}$-space will be 
directed along the $y$-axis according to our remark above. 

 The corresponding asymptotic behavior of $3$-dimensional
conductivity tensor can then be written in the following form
(~\cite{lifazkag}):

\vspace{0.5cm}

 Case 1 (closed orbits):

\begin{equation}
\label{sigcltr}
\sigma^{ik} \simeq {n e^{2} \tau \over m^{*}} \,
\left( \begin{array}{ccc}
(\omega_{B}\tau)^{-2} & (\omega_{B}\tau)^{-1} &
(\omega_{B}\tau)^{-1} \cr
(\omega_{B}\tau)^{-1} & (\omega_{B}\tau)^{-2} &
(\omega_{B}\tau)^{-1} \cr
(\omega_{B}\tau)^{-1} & (\omega_{B}\tau)^{-1} & *
\end{array} \right)
\end{equation}

 Case 2 (open periodic orbits):

\begin{equation}
\label{sigoptr}
\sigma^{ik} \simeq {n e^{2} \tau \over m^{*}} \,
\left( \begin{array}{ccc}
(\omega_{B}\tau)^{-2} & (\omega_{B}\tau)^{-1} &
(\omega_{B}\tau)^{-1} \cr
(\omega_{B}\tau)^{-1} & * & * \cr
(\omega_{B}\tau)^{-1} & * & *
\end{array} \right)
\end{equation}

 Here $\simeq$ means "of the same order in $\omega_{B}\tau$
and $*$ are some constants $\sim 1$. Let us mention also that
the relations (\ref{sigcltr})-(\ref{sigoptr}) give only the
absolute values of $\sigma^{ik}$.

 More complicated types of open electron orbits were constructed
in ~\cite{lifpes1}-~\cite{lifpes2}. Let represent here these
results in a brief form.

 "Thin spatial net". 

 The form of the thin spatial net is shown on the Fig. 3,a.
The net corresponds to the cubic symmetry of the crystal and the 
thickness of tubes is considerably smaller than the periods of the
net.  
 
\begin{figure}
  \begin{picture}(0, 130)
    \leavevmode
    \put(0,0){\epsfxsize=1.0\hsize \epsfbox{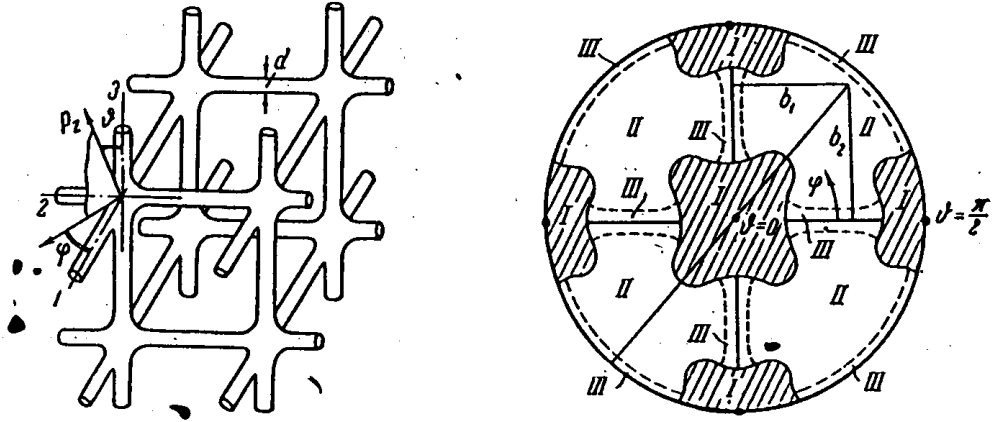}}
  \end{picture}
  \caption{The "thin spatial net" and the corresponding zones on the
unit sphere where the open trajectories exist. As was observed in
~\cite{lifpes1} the mean directions of open orbits are given by the
intersections of planes orthogonal to ${\bf B}$ with the coordinate
planes $xy$, $yz$, $xz$. }
\end{figure}

 We can parameterize now the directions of ${\bf B}$ by the points of 
the unit sphere $S^{2}$ and try to find those directions for which 
we have the open electron orbits on the net. As was pointed out in
~\cite{lifpes1} the open electron orbits exist in this case only for 
six small regions on the unit sphere close to the main 
crystallographic directions 
$(\pm 1,0,0)$, $(0,\pm 1,0)$ and $(0,0,\pm 1)$
(see Fig. 3b). For the directions of ${\bf B}$ lying out of these
domains the open electron orbits do not appear. Let us mention
that this type of trajectories is obviously different from that 
shown on Fig. 2b. This circumstance can be easily seen from the fact 
that in the case of periodic ("warped") cylinder the open 
trajectories exist only for the directions of ${\bf B}$ orthogonal
to the axis of cylinder and do not appear for any other direction.
In the case of thin spatial net we now have the whole regions on the 
unit sphere corresponding to non-closed orbits. These new 
trajectories are not periodic anymore. However, it was shown in
~\cite{lifpes1} that they all have the mean asymptotic directions
given by the intersections of the corresponding planes 
$\Pi({\bf B})$ orthogonal to ${\bf B}$ with the coordinate planes
$xy$, $yz$ and $xz$ (orthogonal to the corresponding main
crystallographic directions). Let us pay here the special
attention to the last two circumstances. We will come back to these
facts when discuss the general topological approach to the
classification problem.

 As was stated in \cite{lifpes1} the corresponding contribution 
to the conductivity can be also written for such orbits in the form
(\ref{sigoptr}) with the $x$-axis directed along the asymptotic
direction of trajectories in ${\bf p}$-space.

 Now let us represent also the results of ~\cite{lifpes2}
concerning the analytical Fermi surfaces given by the 
finite-parametric family of the form:

$$\alpha \left( cos {ap_{x} \over \hbar} + cos {ap_{y} \over \hbar} 
+ cos {ap_{z} \over \hbar} \right) + $$

$$+ \beta 
\left( cos {ap_{x} \over \hbar} \, cos {ap_{y} \over \hbar} +
cos {ap_{y} \over \hbar} \, cos {ap_{z} \over \hbar} +
cos {ap_{x} \over \hbar} \, cos {ap_{z} \over \hbar} \right) +$$

\begin{equation}
\label{anferms}
+ \delta \, cos {ap_{x} \over \hbar} \, cos {ap_{y} \over \hbar} 
\, cos {ap_{z} \over \hbar} \, = \, \zeta_{0}
\end{equation}

 The form of the Fermi surface will now depend on the values of the
parameters $\alpha$, $\beta$, $\delta$ and $\zeta_{0}$. As was
shown in ~\cite{lifpes2} the open electron orbits exist in this
case for four additional different topological types of the Fermi 
surfaces given by (\ref{anferms}) (excluding the spatial net
described above). According to ~\cite{lifpes2} the
open orbits exist in these cases in open zones 
and on the one-dimensional curves on the unit sphere
represented on the Fig. 4.

\begin{figure}
  \begin{picture}(0, 280)
    \leavevmode
    \put(0,0){\epsfxsize=1.0\hsize \epsfbox{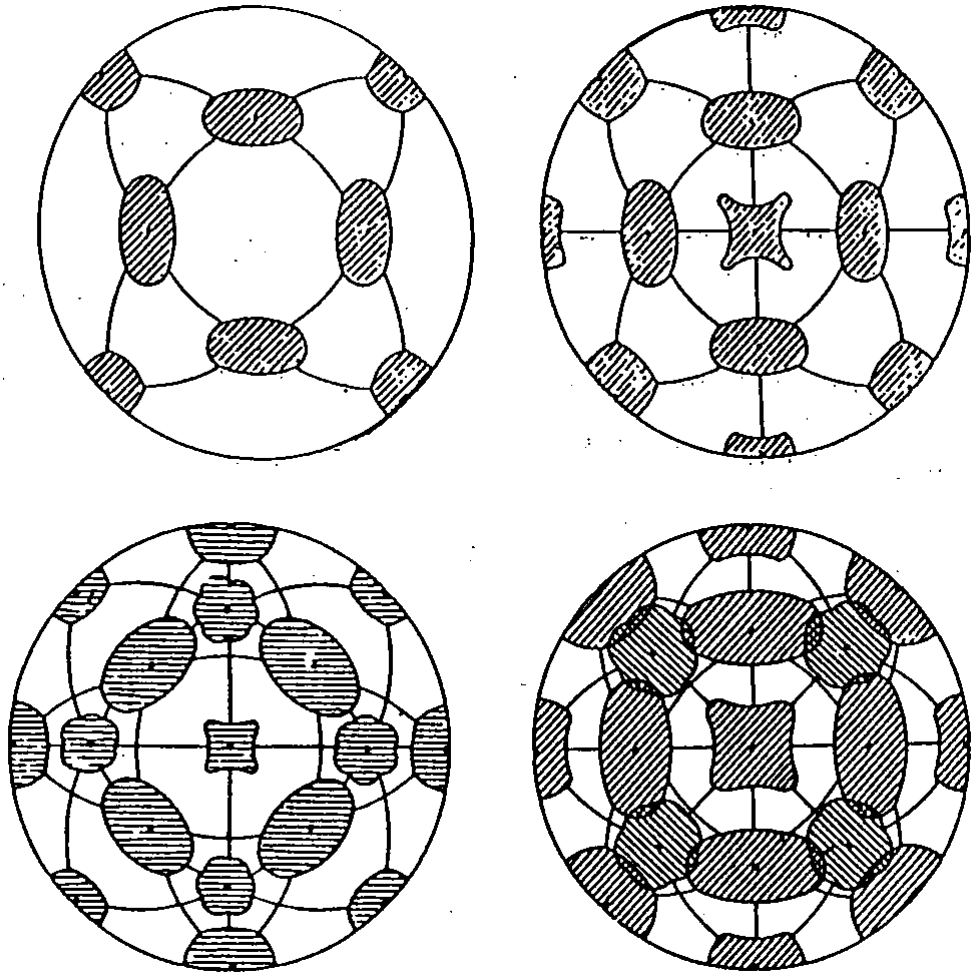}}
  \end{picture}
  \caption{The open zones and the one-dimensional circles
on the unit sphere corresponding to 
open orbits for four different Fermi surfaces from
the family (\ref{anferms}) represented in ~\cite{lifpes2}.
The last picture contains a conceptual mistake contradicting 
to Topological Resonance phenomenon. Namely, the angle
diagram can not contain the whole open domains on the
unit sphere where the open trajectories with different
mean direction exist.}
\end{figure}

 We will discuss later the general topological classification
of arbitrary complicated Fermi surfaces. Let us just make some
comments about the picture on Fig. 4. The dark domains 
on the Fig. 4 give the positions of the largest zones
corresponding to open orbits for the given topological types
of Fermi surface. These domains always exist for the Fermi
surfaces described above being connected with the symmetry
of dispersion relation and the topology of the Fermi surface.
However, as we will discuss later, without any requirements
on the "thickness of tubes" the topological
form of the Fermi surface itself does not determine all the 
regions on the unit sphere corresponding to the open electron
orbits. So, in all the cases on Fig. 4 we predict also an infinite
number of smaller zones on the unit sphere corresponding to
open orbits which forms and positions will depend on the
parameters $\alpha$, $\beta$, $\delta$ and $\zeta_{0}$. Besides
that, there will be some special "unstable" points on the unit 
sphere where the very complicated open orbits can exist being 
completely unstable w.r.t. the small rotations of the magnetic 
field and change of parameters $\alpha$, $\beta$, $\delta$, 
$\zeta_{0}$. 

 Let us also point out here the mistake on the last picture 
of Fig. 4 connected with the overlapping of the domains 
corresponding to open orbits. It was claimed in \cite{lifpes2}
that the open regions with different mean directions exist in the 
intersections of dark domains on the last picture. However,
this situation contradicts to the Topological Resonance phenomenon
which we will discuss below. As we will see then these domains 
can not intersect each other over the whole open regions on the 
unit sphere. So we claim that the dark regions on the last picture  
should be actually smaller and do not overlap each other.

 Let us consider now the general topological approach to the
classification of the open orbits for arbitrary smooth 
three-periodic Fermi surfaces in $R^{3}$.

 The general problem of classification of quasiclassical electron
orbits for arbitrary Fermi surface was set by S.P.Novikov in 
~\cite{nov1}. This problem has been studied by his pupils since
1980's. The important contribution was made by
A.V.Zorich, I.A.Dynnikov and S.P.Tsarev. During 
this period the deep topological results were obtained which 
form now the modern understanding of situation. The general
picture is rather non-trivial and includes the generic behavior
and the special degenerate cases. During the last years the
valuable numerical calculations were done also by R.D.Leo
(\cite{DeLeo}).

 In particular, using the topological resonance following from 
the proofs of these theorems the present authors invented 
the so-called "Topological quantum numbers" 
observable in the conductivity of normal metals 
(\cite{novmal1}). 
These characteristics are represented by the integral planes 
connected with zones on the unit sphere corresponding 
to open orbits. The total set of such planes together with 
the geometry of corresponding zones gives the important topological
characteristic of the dispersion relation in metal.

 It was shown also (S.P.Tsarev, I.A.Dynnikov) that the 
chaotic open orbits can also exist and reveal much more
complicated behavior. We will discuss later these 
cases in details.

 Before going further let us introduce the basic definitions for
the complicated Fermi surfaces in $R^{3}$.

\begin{de}

\vspace{0.5cm}

I) Genus.

 Let us now come back to the original phase space 
$T^{3} = R^{3}/\Gamma^{*}$. The reciprocal lattice 
$\Gamma^{*}$ is generated
by the vectors ${\bf g}_{1}$, ${\bf g}_{2}$, ${\bf g}_{3}$
connected with the vectors ${\bf l}_{1}$, ${\bf l}_{2}$, 
${\bf l}_{3}$ of the physical lattice $\Gamma$ by the simple
formulas:

$${\bf g}_{1} = 2\pi\hbar \, {{\bf l}_{2} \times {\bf l}_{3} \over 
({\bf l}_{1}, {\bf l}_{2}, {\bf l}_{3}) } \,\,\, , \,\,\,
{\bf g}_{2} = 2\pi\hbar \, {{\bf l}_{3} \times {\bf l}_{1} \over  
({\bf l}_{1}, {\bf l}_{2}, {\bf l}_{3}) } \,\,\, , \,\,\,
{\bf g}_{3} = 2\pi\hbar \, {{\bf l}_{1} \times {\bf l}_{2} \over  
({\bf l}_{1}, {\bf l}_{2}, {\bf l}_{3}) } $$

 Every component of the Fermi surface becomes then the smooth 
orientable 2-dimensional surface embedded in $T^{3}$. We can then
introduce the standard genus of every component of the Fermi
surface $g = 0,1,2,...$ according to standard topological
classification depending on if this component is topological
sphere, torus, sphere with two holes, etc ... .

\vspace{0.5cm}

II) Topological Rank.

 Let us introduce the Topological Rank $r$ as the characteristic
of the embedding of the Fermi surface in $T^{3}$. It's much more 
convenient in this case to come back to the total ${\bf p}$-space
and consider the connected components of the three-periodic 
surface in $R^{3}$.

1) The Fermi surface has Rank $0$ if every its connected component
can be bounded by a sphere of finite radius.

2) The Fermi surface has Rank $1$ if every its connected component
can be bounded by the periodic cylinder of finite radius and there
are components which can not be bounded by the sphere.

3) The Fermi surface has Rank $2$ if every its connected component
can be bounded by two parallel (integral) planes in $R^{3}$ and
there are components which can not be bounded by cylinder.

4) The Fermi surface has Rank $3$ if it contains components which
can not be bounded by two parallel planes in $R^{3}$.
\end{de}

 The pictures on Fig. 5, a-d represent the pieces of the Fermi 
surfaces in $R^{3}$ with the Topological Ranks $0$, $1$, $2$ and
$3$ respectively.

\begin{figure}
  \begin{picture}(0, 210)
    \leavevmode
    \put(0,0){\epsfxsize=1.0\hsize \epsfbox{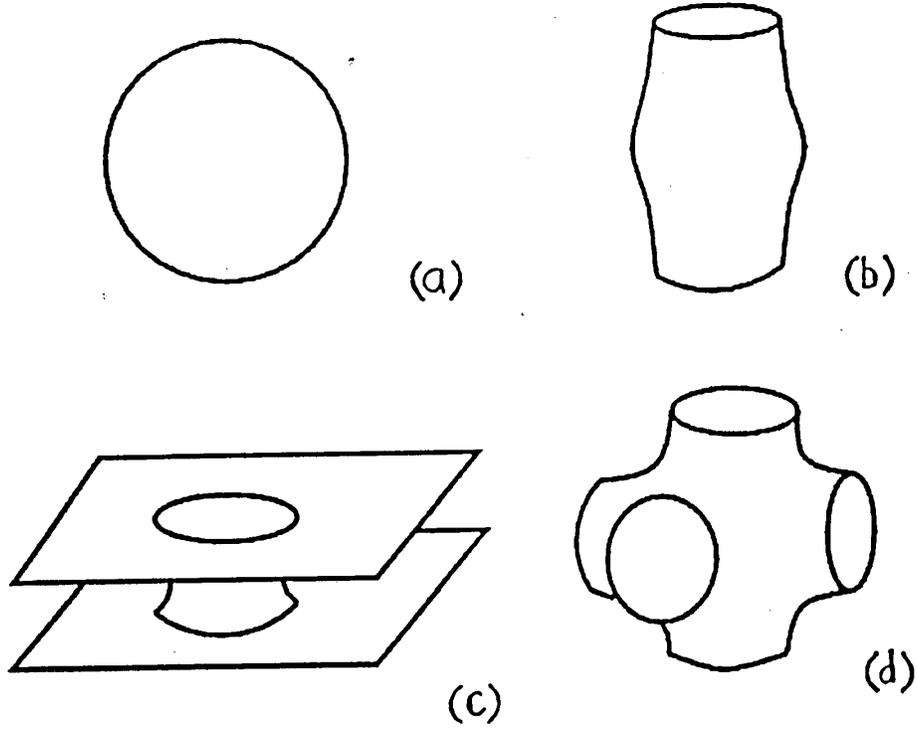}}
  \end{picture}
  \caption{The Fermi surfaces with Topological Ranks $0$, $1$, $2$ 
 and $3$ respectively.}
\end{figure}

 It is easy to see also that the topological Rank coincides with the
maximal Rank of the image of mapping 
$\pi_{1}(S^{i}) \rightarrow \pi_{1}(T^{3})$ for all the connected
components of the Fermi surface.

 As can be seen the genuses of the surfaces
represented on the Fig. 5, a-d are also equal to $0$, $1$, $2$ and
$3$ respectively. However, the genus and the Topological Rank are not 
necessary equal to each other in the general situation. 

 Let us discuss briefly the connection between the genus and the 
Topological Rank since this will play the crucial role in further
consideration. It is easy to see that the Topological Rank of the 
sphere can be only zero and the Fermi surface consists in this case 
of the infinite set of the periodically repeated spheres $S^{2}$
in $R^{3}$. 

 The Topological Rank of the torus $T^{2}$ can take
three values $r = 0$, $r = 1$ and $r = 2$. 

 It is easy to see that all the three cases of periodically 
repeated tori $T^{2}$ in 
$R^{3}$, periodically repeated  "warped" integral 
cylinders and the periodically repeated "warped"
integral planes give the topological 2-dimensional
tori $T^{2}$ in $T^{3}$ after the factorization. (Let us note
here that we call the cylinder in $R^{3}$ integral if it's axis
is parallel to some vector of the reciprocal lattice, while the
plane in $R^{3}$ is called integral if it is generated by some 
two reciprocal lattice vectors.) The case $r=2$, however, has an
important difference from the cases $r=0$ and $r=1$. The matter is 
that the plane in $R^{3}$ is not homological to zero in $T^{3}$
(i.e. does not restrict any domain of "lower energies") after
the factorization. We can conclude so that if these plains
appear as the connected components of the physical Fermi surface
they should always come in pairs, $\Pi_{+}$ and $\Pi_{-}$,
which are parallel to each other in $R^{3}$. The factorization
of $\Pi_{+}$ and $\Pi_{-}$ gives then the two tori 
$T^{2}_{+}$, $T^{2}_{-}$ with the opposite homological classes
in $T^{3}$ after the factorization. The space between the
$\Pi_{+}$ and $\Pi_{-}$ in $R^{3}$ can now be taken as the domain
of lower (or higher) energies and the disjoint union 
$\Pi_{+} \cup \Pi_{-}$ will correspond to the union 
$T^{2}_{+} \cup T^{2}_{-}$ homological to zero in $T^{3}$.

 It can be shown that the Topological Rank of any component of 
genus $2$ can not exceed $2$ also. The example of the corresponding 
immersion of such component with maximal Rank is shown at Fig. 5, c 
and represents the two parallel planes connected by cylinders.

 At last we say that the Topological Rank of the components with
genus $g \geq 3$ can take any value $r = 0,1,2,3$.

 Let us also show at last two "exotic examples" of the Fermi surfaces
of Rank 1 and 2 respectively (see Fig. 6, a,b).

\begin{figure}
  \begin{picture}(0, 170)
    \leavevmode
    \put(0,0){\epsfxsize=1.0\hsize \epsfbox{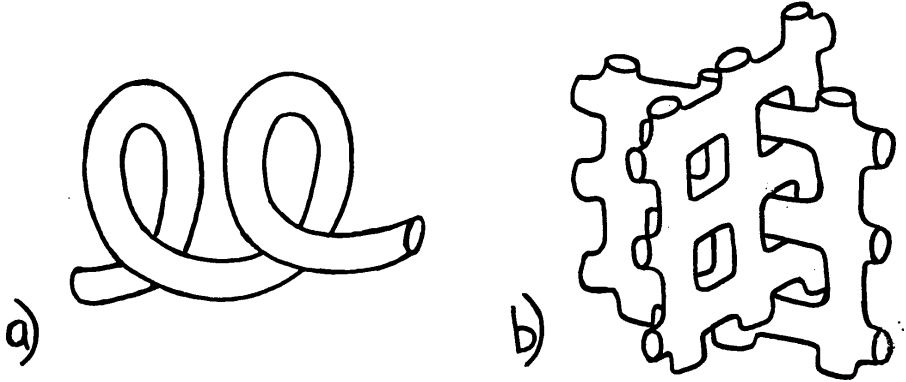}}
  \end{picture}
  \caption{(a) Connected component of Rank $1$ having the form of 
"helix". Open orbits are absent for any direction of ${\bf B}$.
(b) The example of the Fermi surface of Rank $2$ containing two 
components with different integral directions.}
\end{figure}

 We are going to formulate now the topological theorems concerning
the general situation of any complicated Fermi surfaces. We will
assume now that the dispersion relation $\epsilon({\bf p})$ is a 
Morse function on $T^{3}$ and consider the non-singular energy
levels $\epsilon({\bf p}) = const$ such that 
$\nabla \epsilon({\bf p}) \neq 0$ everywhere on the corresponding
surface. It is easy to see that all the reconstructions of the
constant energy surface take place only at the points of 
singularity. Such the topological type (genus) and the Topological 
Rank of the constant energy surface are constant on the intervals
of regularity. The number of singular constant energy levels
is finite for the Morse function $\epsilon({\bf p})$.

 The electron trajectories will now be given by the intersections
of constant energy surfaces with the planes orthogonal to the 
magnetic field ${\bf B}$. At every plane $\Pi$ orthogonal to 
${\bf B}$  they can be then considered as the level curves of the
quasiperiodic function 
${\hat \epsilon}({\bf p}) = \epsilon({\bf p})|_{\Pi}$ with three 
quasiperiods. For the case of purely rational directions of 
${\bf B}$ the corresponding functions become purely periodic.
The trajectories can also be represented as the level curves
of the height function 
$h({\bf p}) = B_{1} p_{1} + B_{2} p_{2} + B_{3} p_{3}$ restricted
on the constant energy levels. This function, however, is not
uniquely defined in the three-dimensional torus 
$T^{3} = R^{3}/\Gamma^{*}$ and becomes the $1$-form in $T^{3}$ 
after the compactification. The corresponding electron trajectories 
become  then the level curves of the $1$-form in $T^{3}$ restricted 
to the compact smooth energy levels $\epsilon({\bf p}) = c$.

 We will assume now that the restrictions ${\hat h}({\bf p})$
of $h({\bf p})$ on the constant energy surfaces in $R^{3}$
give also the Morse functions, i.e. all the critical points where
$\nabla \epsilon({\bf p}) \parallel {\bf B}$ are non-degenerate
on these surfaces.

\begin{de}
 We call the electron trajectory non-singular if 
it is not adjacent to the critical point of ${\hat h}({\bf p})$.
The trajectories adjacent to critical points of 
${\hat h}({\bf p})$ (and the critical points themselves) we
will call singular.
\end{de}

 According to our assumption there are three types of the critical
points of ${\hat h}({\bf p})$ on the constant energy levels.
Namely, we can have the local minimum, the saddle point and the 
local maximum in the ${\bf p}$-space (Fig. 7, a-c).

\begin{figure}
  \begin{picture}(0, 170)
    \leavevmode
    \put(0,0){\epsfxsize=1.0\hsize \epsfbox{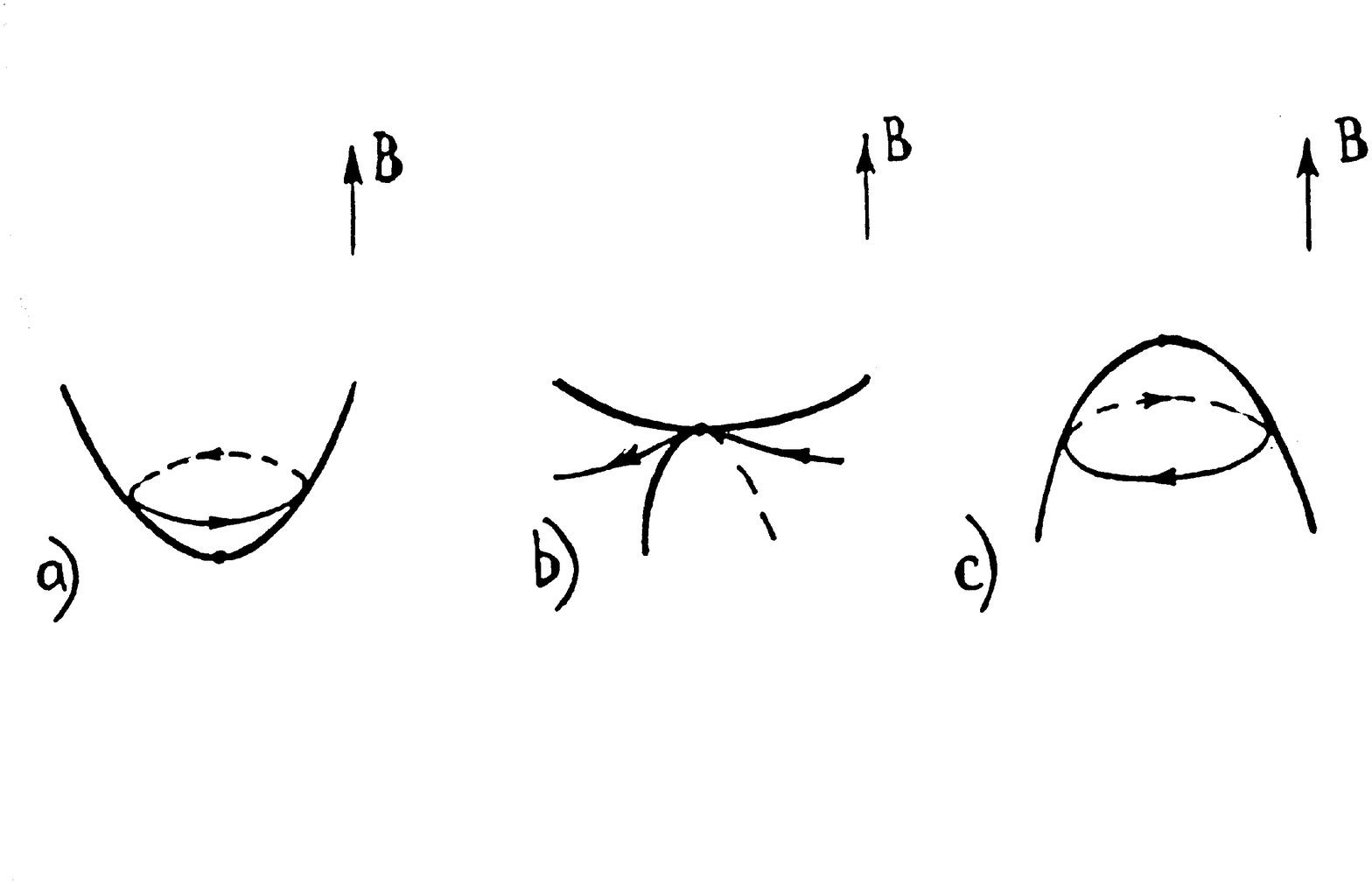}}
  \end{picture}
  \caption{The critical points of the function ${\hat h}({\bf p})$
and the corresponding singular trajectories.}
\end{figure}

 The corresponding pictures in the planes orthogonal to ${\bf B}$
are represented on the Fig. 8, a-c.

\begin{figure}
  \begin{picture}(0, 170)
    \leavevmode
    \put(0,0){\epsfxsize=1.0\hsize \epsfbox{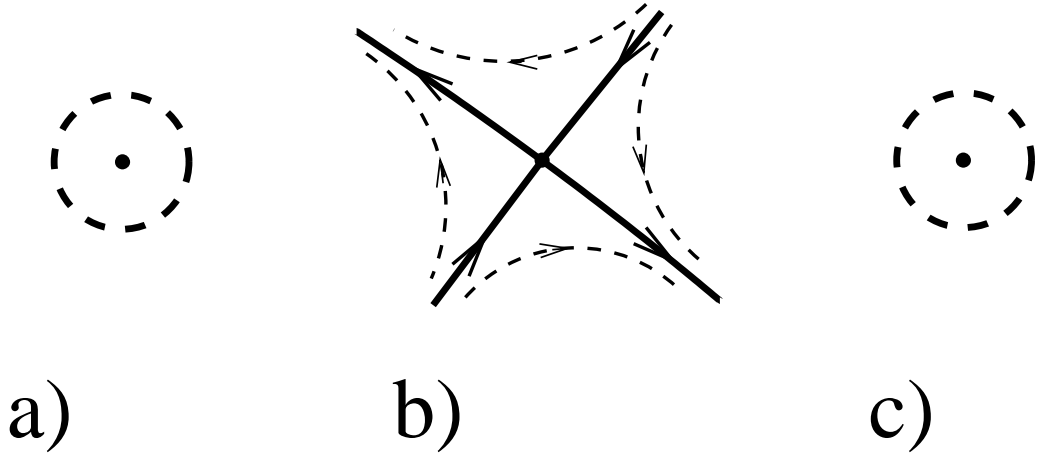}}
  \end{picture}
  \caption{The singular trajectories in the plane
  orthogonal to ${\bf B}$.}
\end{figure}

 Let us give also the definitions of "rationality" and 
"irrationality" of the direction of ${\bf B}$.

\begin{de}
Let $\{{\bf g}_{1}, {\bf g}_{2}, {\bf g}_{3}\}$ be the basis of the
reciprocal lattice $\Gamma^{*}$. Then:

 1) The direction of ${\bf B}$ is rational (or has irrationality $1$)
if the numbers $({\bf B}, {\bf g}_{1})$, $({\bf B}, {\bf g}_{2})$,
$({\bf B}, {\bf g}_{3})$ are proportional to each other with
rational coefficients.

 2) The direction of ${\bf B}$ has irrationality $2$ if the numbers
$({\bf B}, {\bf g}_{1})$, $({\bf B}, {\bf g}_{2})$, 
$({\bf B}, {\bf g}_{3})$ generate the linear space of dimension
$2$ over $Q$.

 3) The direction of ${\bf B}$ has irrationality $3$ if the numbers
$({\bf B}, {\bf g}_{1})$, $({\bf B}, {\bf g}_{2})$,
$({\bf B}, {\bf g}_{3})$ are linearly independent over $Q$.
\end{de} 

 The conditions (1)-(3) can be formulated also as if the plane
$\Pi({\bf B})$ orthogonal to ${\bf B}$ contains two linearly
independent reciprocal lattice vectors, just one linearly
independent reciprocal lattice vector or no reciprocal lattice
vectors at all respectively.

 It can be seen also that if 
$\{{\bf l}_{1}, {\bf l}_{2}, {\bf l}_{3}\}$ is the basis of
the original lattice in ${\bf x}$-space then the irrationality of
the direction of ${\bf B}$ will be given by the dimension of the
vector space generated by numbers

$$({\bf B}, {\bf l}_{2}, {\bf l}_{3}) \,\,\, , \,\,\,
({\bf B}, {\bf l}_{3}, {\bf l}_{1}) \,\,\, , \,\,\,
({\bf B}, {\bf l}_{1}, {\bf l}_{2}) $$
over $Q$. Easy to see that these numbers have the meanings of the
the magnetic fluxes through the faces of elementary lattice cell.
In all our considerations they will always be much smaller than 
the quantum of magnetic flux and their absolute values will
not be important for the quasiclassical pictures. However, their
ratios having the pure geometrical meanings will play the important
role as we will see later.

 We are going to consider now the geometry of the non-singular 
electron trajectories. Let us start with the simplest cases.

 1) The Fermi surface has Topological Rank $0$. 

 All the components
of the Fermi surface are compact in $R^{3}$ in this case and there 
is no open trajectories at all.

 2) The Fermi surface has Topological Rank $1$. 

 In this case we
can have both open and closed electron trajectories. However
the open trajectories (if they exist) should be quite simple in 
this case. They can arise only if the magnetic field is orthogonal
to the mean direction of one of the components of Rank $1$ and
are periodic with the same integer mean direction. There is only
the finite number of possible mean directions of open orbits in
this case and a finite "net" of one-dimensional curves on the
unit sphere giving the directions of ${\bf B}$ corresponding
to the open orbits. In some special points we can have the 
trajectories with different mean directions lying in different
parallel planes orthogonal to ${\bf B}$. Easy to see that in this 
case the direction of ${\bf B}$ should be purely rational such
that the orthogonal plane $\Pi({\bf B})$ contains two different
reciprocal lattice vectors. It is evident also that there is only 
the finite number of such directions of 
${\bf B}$ clearly determined
by the mean directions of the components of Rank $1$. Let us 
mention also that the existence 
of open orbits is not necessary here
even for ${\bf B}$ orthogonal to the mean direction of some 
component of Rank $1$ as can be seen from the example of the
"helix" represented on Fig. 6, a.

 3) The Fermi surface has Topological Rank 2.

 It can be easily seen that this case gives much more 
possibilities for the existence of open orbits for different
directions of the magnetic field. In particular, this is the
first case where the open orbits can exist for the generic 
direction of ${\bf B}$ with irrationality $3$. So, in this case
we can have the whole regions on the unit sphere such that the 
open orbits present for any direction of 
${\bf B}$ belonging to the
corresponding region. It is easy to see, however, that the open
orbits have also a quite simple description in this case.
Namely, any open orbit (if they exist) lies in this case in
the straight strip of the finite width for any direction of
${\bf B}$ not orthogonal to the integral planes given by the
components of Rank $2$. The boundaries of the corresponding strips
in the planes $\Pi({\bf B})$ orthogonal to ${\bf B}$ will be 
given by the intersection of $\Pi({\bf B})$ with the pairs of
integral planes bounding the corresponding components of Rank $2$.
It can be also shown (\cite{dynn1}, \cite{dynn2}) that every open 
orbit passes through the strip from $- \infty$ to  $+ \infty$
and can not turn back. 

 The contribution of every family of orbits 
with the same direction to the conductivity coincides in this
case with the formula (\ref{sigoptr}) and reveals the strong
anisotropy when $\omega_{B}\tau \rightarrow \infty$.

 For purely rational directions of ${\bf B}$ we can have the 
situation when the open trajectories with different mean
directions present on different components of the Fermi surface.
For example, for the "exotic" surface shown at Fig. 6, b
we will have the periodic trajectories along both the $x$ and $y$
directions in different planes orthogonal to ${\bf B}$ if 
${\bf B}$ is directed along the $z$ axis. However, it can be 
shown that for any direction of ${\bf B}$ which is not purely
rational this situation is impossible. We have so that for any
direction of ${\bf B}$ with irrationality $2$ or $3$ all the
open orbits will have the same mean direction and can exist
only on the components of Rank $2$ with the same (parallel) 
integral orientation. This statement is a corollary 
of more general topological theorem which we will discuss 
in the next part.

 At last we note that the directions of ${\bf B}$ orthogonal
to one of the components of Rank $2$ are purely rational and
all the non-singular open orbits (if they exist) are rational
periodic in this case. 
For any family of such orbits with the same mean direction
the corresponding contribution to the
conductivity can then be written in the form (\ref{sigoptr})
in the appropriate coordinate system. However, the direction 
of open orbits can not be predicted apriori
in this case. We will discuss these questions later when
consider the "Special rational directions" for the case
of arbitrary Fermi surfaces.

 Let us discuss now the most general and complicated case of
the arbitrary Fermi surface of Topological Rank $3$. 
We describe first the convenient procedure 
(\cite{dynn4},\cite{dynn7}) of reconstruction of the constant
energy surface when the direction of ${\bf B}$ is fixed.

 Let us fix the direction of ${\bf B}$ 
and consider all closed (in $R^{3}$) non-singular electron 
trajectories on the given energy level. The parts of the constant
energy surface covered by the non-singular closed trajectories
can be either the tori or the cylinders in $R^{3}$ bounded by the
singular trajectories (some of them maybe just points of minimum
or maximum) at the bottom and at the top (see Fig. 9).

\begin{figure}
  \begin{picture}(0, 180)
    \leavevmode
    \put(0,0){\epsfxsize=1.0\hsize \epsfbox{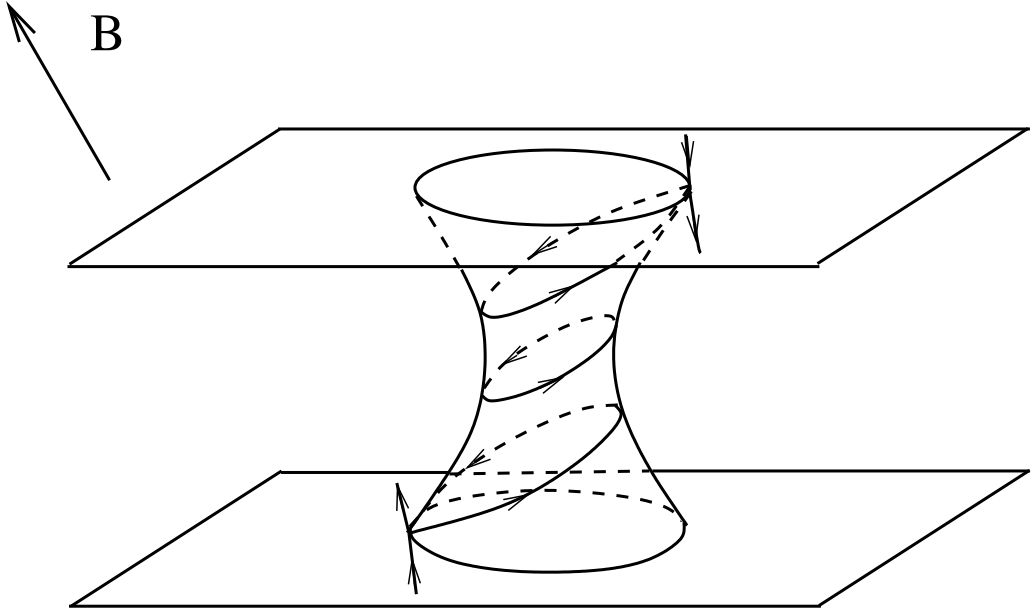}}
  \end{picture}
\caption{The cylinder of closed trajectories bounded by the singular
orbits. (The simplest case of just one critical point on the singular
trajectory.)}
\end{figure}

 Let us remove from the Fermi surface the parts containing the 
compact nonsingular trajectories. The remaining part

$$S_{F}/({\rm Compact \, Nonsingular \, Trajectories}) \,\,\, =
\,\,\, \cup_{j} \, S_{j}$$
is a union of the $2$-manifolds $S_{j}$ with boundaries 
$\partial S_{j}$ who are the compact singular trajectories.
The generic type is a separatrix orbit with just one 
critical point like on the Fig. 10.

\begin{de}
We call every piece $S_{j}$ the 
{\bf "Carrier of open trajectories"}. The trajectory is "chaotic"
if the genus $g(S_{j})$ is greater than $1$. The case
$g(S_{j}) = 1$ we call "Topologically Completely Integrable".
\footnote{Such systems on $T^{2}$ were discussed for example
in \cite{arnold}; the generic open orbits 
are topologically equivalent to the straight
lines. Ergodic properties of such systems indeed can be nontrivial
as it was found by Ya.Sinai and K.Khanin in \cite{sinkhan}.}
\end{de}

 Let us fill in the holes by 
topological $2D$ discs lying in the planes orthogonal to ${\bf B}$
and get the closed surfaces  

$${\bar S}_{j} \,\,\, = \,\,\, S_{j} \cup (2-discs)$$
(see Fig. 10).

\begin{figure}
  \begin{picture}(0, 200)
    \leavevmode  
 \put(0,0){\epsfxsize=0.9\hsize \centerline{\epsfbox{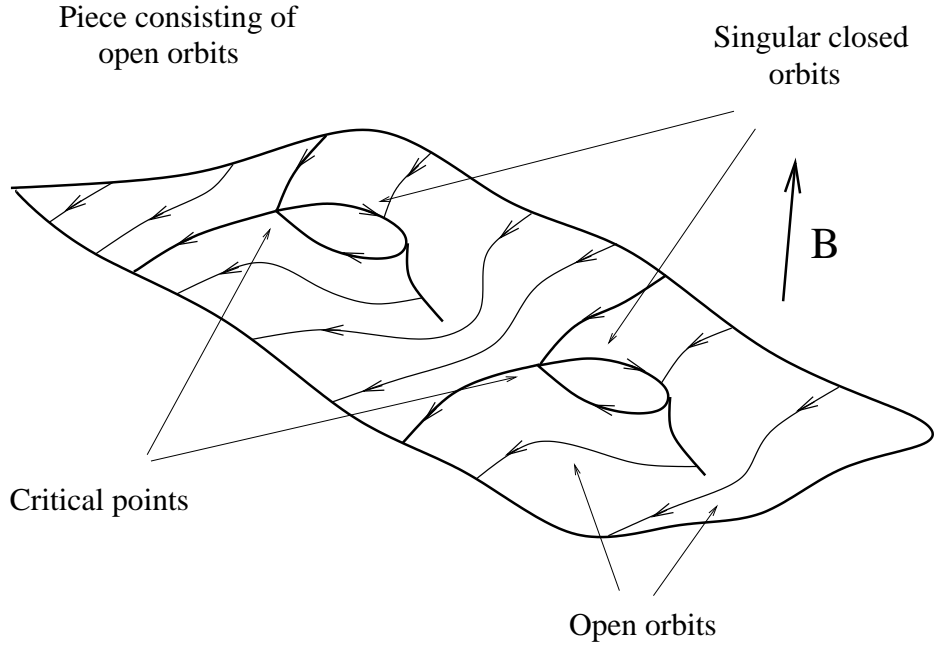}}}
  \end{picture}
\caption{The reconstructed constant energy surface 
with removed compact orbits and the two-dimensional discs attached 
to the singular orbits in the generic case of just one critical 
point on every singular orbit. }
\end{figure}

 This procedure gives again the periodic surface 
${\bar S}_{\epsilon}$ after the reconstruction and we can define 
the "compactified carriers of open trajectories" both in $R^{3}$ 
and $T^{3}$. 

 Let us formulate now the main topological theorems concerning 
the geometry of open trajectories which made a breakthrough
in the theory of such dynamical systems on the Fermi surfaces
(\cite{zorich}, \cite{dynn3}).

{\bf Theorem $1$.} \cite{zorich}

{\it Let us fix the energy level $S_{\epsilon}$ and any rational
direction ${\bf B}_{0}$ such that no two saddle points on
$S_{\epsilon}$ are connected in $R^{3}$ by the singular electron
trajectory. Then for all the directions of ${\bf B}$ 
close enough to ${\bf B}_{0}$ every open
trajectory lies in the strip of the finite width between two
parallel lines in the plane orthogonal to ${\bf B}$.
}

 In fact, the proof of the Theorem $1$ was based on 
the statement that genus of every compactified carrier of open
orbits ${\bar S}_{j}$ is equal to $1$ in this case.

{\bf Theorem $2$.} \cite{dynn3}

{\it Let a generic dispersion relation

$$\epsilon({\bf p}): \,\,\, T^{3} \,\,\, \rightarrow R$$
be given such that for level 
$\epsilon({\bf p}) = \epsilon_{0}$ the genus $g$ of some
carrier of open trajectories ${\bar S}_{i}$ is greater than $1$.
Then there exists an open interval 
$(\epsilon_{1}, \epsilon_{2})$ containing $\epsilon_{0}$
such that for all $\epsilon \neq \epsilon_{0}$ in this interval
the genus of carrier of open trajectories is less than $g$.
}

 The Theorem $2$ claims then that only the 
"Topologically Completely Integrable case" can be stable with
respect to the small variation of energy level and has the 
generic properties in this situation. For the generic dispersion
relations it follows also from the Theorem 1 that
this case is the only stable case with respect to the 
small rotations of the magnetic field ${\bf B}$ on the
unit sphere $S^{2}$ (see the survey \cite{dynn7} for details).

{\bf Physical results of the present authors (\cite{novmal1},
\cite{novmal2}).}
 
 The very important property of Compactified Carriers of
open orbits in the stable case was pointed out by the present
authors (\cite{novmal1})
and called later the {\bf "Topological Resonance"}. This
property plays the crucial role for the physical phenomena
and was first used in \cite{novmal1} (see also \cite{novmal2})
where the "Topological Quantum Numbers" 
observable in the conductivity were introduced.
Namely, consider the stable "Topologically Completely Integrable 
case" corresponding to genus $1$ of the carriers of open 
trajectories ${\bar S}_{i}$. The "Topological Resonance"
dictated by the elementary differential topology claims that 
all the tori $T^{2}$
represented by ${\bar S}_{i}$ do not intersect each other
and have the same (up to the sign) non-divisible homology
class in $H_{2}(T^{3})$. For the generic (irrationality $3$)
directions of ${\bf B}$ the corresponding coverings 
of ${\bar S}_{i}$ in $R^{3}$ look like the warped planes
with mean directions parallel to the same integral plane in
$R^{3}$; the open trajectories in $R^{3}$
have then the same mean directions in the planes orthogonal to
${\bf B}$.
{\bf Our conclusion is that the corresponding 
contribution of all these trajectories to 
the conductivity tensor has then the same form (\ref{sigoptr})
in the appropriate coordinate system common for all of them.
 This fundamental fact
leads to existence of measurable characteristics having the 
topological origin in the conductivity of normal metals. }

  Let us take a single crystal of a normal metal and consider
the full angle diagram of the conductivity $\sigma^{ik}$
for all the directions of ${\bf B}$ parameterized by the points
of the unit sphere. For the real single-crystal metal
only the orbits close to the Fermi surface will give the   
contribution to the conductivity tensor. 

 Now for all regions where we have just the closed trajectories
on the Fermi surface we will have the asymptotic behavior
(\ref{sigcltr}) of the conductivity tensor as
$B\tau \rightarrow \infty$. The longitudinal conductivity then
remains constant in the direction of ${\bf B}$ and decreases
as $B \rightarrow \infty$ for all the orthogonal directions
of electric field. Any other behavior of conductivity tensor 
shows in this case the presence of open electron trajectories
on the Fermi surface lying in the planes orthogonal to ${\bf B}$.
We know, however, that for any set of open trajectories stable
under the rotations of ${\bf B}$ we should have the situation  
described in the Theorems 1, 2. 
The corresponding conductivity
tensor $\sigma^{ik}$ is given by the formula (\ref{sigoptr})
in this case and has rank $2$ in the limit
$B \rightarrow \infty$. We can claim then
that any open region on $S^{2}$ with the regular stable behavior
of the conductivity different from (\ref{sigcltr}) should
correspond to (\ref{sigoptr}) and contain the only one direction
($\eta({\bf B})$) in the three-dimensional space where the
conductivity decreases as $(\omega_{B}\tau)^{-2}$ as
$B \rightarrow \infty$. It can be seen from the previous 
considerations that this direction should
coincide with the mean directions of the open orbits in
${\bf p}$-space (let us remind that the projection of the   
electron trajectory on $xy$-plane in ${\bf x}$-space can be
obtained by the rotation of the trajectory in ${\bf p}$-space
by $\pi/2$ in the plane orthogonal to ${\bf B}$).
We can extract now from Theorems 1, 2 
that $\eta({\bf B})$
should always belong to some integral plane 
$\Gamma_{\alpha}$ (with
respect to reciprocal lattice) which is the same for the whole 
stability region on the unit sphere and represents the
homology class $c \in H_{2}(T^{3})$ of the stable two-dimensional
tori $T^{2}_{i}$. The stability of this plane w.r.t. the small
rotations of the magnetic field gives then the easy possibility  
to get this characteristic in the experiment.

 The total set of the stability regions $\Omega_{\alpha}$ on the
unit sphere with the corresponding integral planes
$\Gamma_{\alpha}$ was called in $\cite{novmal1})$ the
"Topological Quantum characteristics" of the normal metal.
These quantities have the quantum origin being obtained from
the apriori unknown dispersion relation $\epsilon({\bf p})$
but appear in a purely geometrical way from the geometry of
the Fermi surface.

 The corresponding integral planes $\Gamma_{\alpha}$ can then
be given by three integer numbers
$(n^{1}_{\alpha}, n^{2}_{\alpha}, n^{3}_{\alpha})$
(up to the common multiplier) from the equation
$$n^{1}_{\alpha} [{\bf x}]_{1} + n^{2}_{\alpha} [{\bf x}]_{2} +
n^{3}_{\alpha} [{\bf x}]_{3} = 0$$
where $[{\bf x}]_{i}$ are the coordinates in the basis
$\{{\bf g}_{1}, {\bf g}_{2}, {\bf g}_{3}\}$ of the reciprocal 
lattice, or equivalently

$$n^{1}_{\alpha} ({\bf x}, {\bf l}_{1}) +
n^{2}_{\alpha} ({\bf x}, {\bf l}_{2}) +
n^{3}_{\alpha} ({\bf x}, {\bf l}_{3}) = 0$$   
where $\{{\bf l}_{1}, {\bf l}_{2}, {\bf l}_{3}\}$ is the basis
of the initial lattice in the coordinate space.

The numbers $(n^{1}_{\alpha}, n^{2}_{\alpha}, n^{3}_{\alpha})$
were called the "Topological Quantum numbers" of a dispersion
relation in metal.

 Let us add also that the number of tori $T_{i}^{2}$ being even
can still be different for the different points of stability
zone $\Omega_{\alpha}$. We can then introduce in the general
situation the "sub-boundaries" of the stability zone which are
the piecewise smooth curves inside $\Omega_{\alpha}$ where   
the number of tori generically changes by $2$.
The asymptotic behavior of conductivity will still be
described by the formula (\ref{sigoptr}) in this case
but the dimensionless coefficients will then "jump"
on the sub-boundaries of stability zone. Let us however
mention here that this situation can be observed only for
rather complicated Fermi surfaces.

 As was first shown by S.P.Tsarev (\cite{tsarev}) the
more complicated chaotic open orbits can still exist on
rather complicated Fermi surfaces $S_{F}$. Such, the
example of open trajectory which does not lie in any finite
strip of finite width was constructed. The corresponding
direction of ${\bf B}$ had the irrationality $2$ in this
example and the closure of the open orbit was a "half" of
the surface of genus $3$ separated by the singular
closed trajectory non-homotopic to zero in $T^{3}$. However,
the trajectory had in this case the asymptotic direction
even not being restricted by any straight strip of finite
width in the plane orthogonal to ${\bf B}$.

 As was shown later in ~\cite{dynn4}, ~\cite{dynn6} this situation
always takes place for any chaotic trajectory for the
directions of ${\bf B}$ with irrationality $2$. We have so,  
that for non-generic "partly rational" directions of ${\bf B}$
the chaotic behavior is still not "very complicated"
and resembles some features of stable open electron trajectories.

  The corresponding asymptotic behavior of conductivity should
reveal also the strong anisotropy properties in the plane
orthogonal to ${\bf B}$ although the exact form of
$\sigma^{ik}$ will be slightly different from (\ref{sigoptr}) for
this type of trajectories. By the same reason, the asymptotic
direction of orbit can be measured experimentally in this case
as the direction of lowest longitudinal conductivity in  
$R^{3}$ according to kinetic theory. The measure of the
corresponding set on the unit sphere
is obviously zero for such type of trajectories
being restricted by the measure of directions of irrationality
$2$.

 The more complicated examples of chaotic open orbits were  
constructed in ~\cite{dynn4} for 
the Fermi surface having genus $3$.
The direction of the magnetic field has the irrationality $3$
in this case and the closure of the chaotic trajectory
covers the whole Fermi surface in $T^{3}$. These types of the
open orbits do not have any asymptotic direction in the planes   
orthogonal to ${\bf B}$ and have rather complicated form
"walking everywhere" in these planes. Let us discuss later
this case in more details.

{\bf The recent topological results 
(\cite{dynn4}, \cite{dynn7}).}

 After the works \cite{zorich},
\cite{dynn3}, \cite{novmal1} a systematic investigation
of open orbits was completed by I.A.Dynnikov 
(see \cite{dynn5}-\cite{dynn4}, \cite{dynn7}).
In particular the total picture of different types of the 
open orbits for generic dispersion relations was presented 
(\cite{dynn7}). Let us describe here the corresponding
topological results.

{\bf Theorem $3$} (\cite{dynn4}, \cite{dynn7}).

{\it Let us fix the dispersion relation 
$\epsilon = \epsilon({\bf p})$
and the direction of ${\bf B}$ of irrationality $3$ and consider
all the energy levels for 
$\epsilon_{min} \leq \epsilon \leq \epsilon_{max}$. Then:

 1) The open electron trajectories exist for all the energy values 
$\epsilon$ belonging to the closed connected energy interval
$\epsilon_{1}({\bf B}) \leq \epsilon \leq \epsilon_{2}({\bf B})$
which can degenerate to just one energy level 
$\epsilon_{1}({\bf B}) = \epsilon_{2}({\bf B}) = 
\epsilon_{0}({\bf B})$.

 2) For the case of the non-degenerate energy interval the set of 
compactified carriers of open trajectories 
${\bar S}_{\epsilon}$ is always a disjoint union of 
two-dimensional tori $T^{2}$ in $T^{3}$ for all
$\epsilon_{1}({\bf B}) \leq \epsilon \leq \epsilon_{2}({\bf B})$.
All the tori $T^{2}$ for all the energy levels do not intersect
each other and have the same (up to the sign) indivisible
homology class $c \in H_{2}(T^{3})$, $c \neq 0$.
The number of tori $T^{2}$ is even for every fixed energy level
and the corresponding covering ${\bar S}_{\epsilon}$ in $R^{3}$
is a locally stable family of parallel ("warped") integral planes
$\Pi^{2}_{i} \subset R^{3}$ with common direction given by $c$.
The form of ${\bar S}_{\epsilon}$ described above is locally 
stable with the same homology class $c \in H_{2}(T^{3})$ under 
small rotations of ${\bf B}$.  
All the open electron trajectories at all the energy levels
lie in the strips of finite width with the same direction and
pass through them. The mean direction of the trajectories is given 
by the intersections of planes $\Pi({\bf B})$ with the integral
family $\Pi^{2}_{i}$ for the corresponding "stability zone" on 
the unit sphere. 

 3) The functions $\epsilon_{1}({\bf B})$, $\epsilon_{2}({\bf B})$
defined for the directions of ${\bf B}$ of irrationality $3$ can
be continuated on the unit sphere $S^{2}$ as the piecewise smooth 
functions such that 
$\epsilon_{1}({\bf B}) \geq \epsilon_{2}({\bf B})$ everywhere
on the unit sphere.

 4) For the case of trivial energy interval 
$\epsilon_{1} = \epsilon_{2} = \epsilon_{0}$ the corresponding
open trajectories may be chaotic. Carrier of the chaotic open
trajectory is homologous to zero in $H_{2}(T^{3},Z)$ and has genus
$\geq 3$. For the generic energy level $\epsilon = \epsilon_{0}$
corresponding directions of magnetic fields $\xi \in S^{2}$
leading to the chaotic trajectories belong to the countable union
of the codimension $1$ subsets. Therefore a measure of this set is 
equal to zero on $S^{2}$.
}

 The whole manifold ${\bar S}_{\epsilon}$ is always homologous to
zero in $T^{3}$ and all the two dimensional tori $T^{2}$ can be
always divided in two equal groups $\{T^{2}_{i+}\}$,
$\{T^{2}_{i-}\}$ according to the direction of the electron motion.
As can be proved using Theorem $1$ (\cite{dynn7}) the "stability
zones" form the everywhere dense set on the unit sphere for the 
generic dispersion relations. All the non-closed trajectories
stable under the small rotations of ${\bf B}$ should have thus
the form described above.
 
 All the trajectories described in Theorem $3$  
give the same form of conductivity 
(\ref{sigoptr}) as $B \rightarrow \infty$ in the appropriate
coordinate system (Topological resonance). 

 Let us consider now the non-generic cases of the magnetic
fields of irrationality $1$ and $2$. The Theorem $3$  
should be slightly modified in this case but has the same main
features as in the case of fully irrational magnetic field
(\cite{dynn7}).
Namely, the set of carriers of open trajectories 
${\bar S}_{\epsilon}$ can
contain now the two-dimensional tori $T^{2\prime}_{s}$ having
the zero homology class in $H_{2}(T^{3})$ in addition to the 
family of parallel tori with non-zero homology classes described 
above. The corresponding covering of these components in $R^{3}$ 
are "warped" periodic cylinders and all the open
trajectories belonging to these components are purely periodic.
As it is easy to see these components of ${\bar S}_{\epsilon}$ 
are stable with respect to the small rotations
of ${\bf B}$ in the plane orthogonal to the axis of cylinder 
and disappear after any other small rotation. The part (1)
of the Theorem 2 will be true also for rational or "partly
rational" directions of ${\bf B}$ with some connected energy
interval 
$\epsilon_{1}^{\prime}({\bf B}) \leq \epsilon \leq 
\epsilon_{2}^{\prime}({\bf B})$. However, the boundary values
$\epsilon_{1}^{\prime}({\bf B})$, $\epsilon_{2}^{\prime}({\bf B})$
do not necessarily coincide in this case with the values of 
piecewise smooth
functions $\epsilon_{1}({\bf B})$, $\epsilon_{2}({\bf B})$ defined
everywhere on $S^{2}$ according to Theorem $3$ (\cite{dynn7}).
Namely, we will have instead the relations
$\epsilon_{1}^{\prime}({\bf B}) \leq \epsilon_{1}({\bf B}) \leq
\epsilon_{2}({\bf B}) \leq \epsilon_{2}^{\prime}({\bf B})$
for all such directions of ${\bf B}$ where all the components
of ${\bar S}_{\epsilon}$ belonging to intervals
$[\epsilon_{1}^{\prime}({\bf B}), \epsilon_{1}({\bf B}))$,
$(\epsilon_{2}({\bf B}), \epsilon_{2}^{\prime}({\bf B})]$
consist of the tori homologous to
zero in $T^{3}$. As we will see this cases can be observed 
experimentally for those ${\bf B}$ where the 
Fermi level lies in the one of such intervals and only the 
"partly stable" non-closed trajectories exist on the Fermi surface.

 The chaotic open orbits can exist also for the directions of
${\bf B}$ of irrationality $2$ (Tsarev chaotic orbits). Their
behavior however reveals completely different properties
from the case of generic directions of magnetic field as we
already discussed above.

 The "partly-stable" cylinders
described above do not intersect the "absolutely stable"
components of ${\bar S}_{\epsilon}$ and all the open
trajectories will still have the same mean direction 
if ${\bf B}$ is not orthogonal to corresponding integral plane
$\Gamma_{\alpha}$ (let us mention that all the trajectories lying
on "regular" parallel integral planes in 
${\bar S}_{\epsilon} \subset R^{3}$ will be also
periodic with the same period in this case).
The form of the conductivity tensor will still be described by
the formula (\ref{sigoptr}) 
but the numerical values of dimensionless coefficients will jump 
for those non-generic directions of ${\bf B}$ where the situation 
described appears. In particular, the asymptotic behavior
(\ref{sigoptr}) will arise on the one-dimensional curves for
those directions of ${\bf B}$ where
$\epsilon_{F} \in 
[\epsilon_{1}^{\prime}({\bf B}), \epsilon_{1}({\bf B})) \cup
(\epsilon_{2}({\bf B}), \epsilon_{2}^{\prime}({\bf B})]$
being stable only for rotations of ${\bf B}$ in the corresponding
direction. As follows from the statements above
the corresponding directions of ${\bf B}$ can have at most
the irrationality $2$ and the corresponding one-dimensional
curves are always the parts of the circles orthogonal to some
integer vector in the reciprocal lattice $\Gamma^{*}$.

 Let us make now a special remark about the "Special directions" 
of ${\bf B}$ orthogonal to the integral planes $\Gamma_{\alpha}$
if this direction belongs to the corresponding stability
zone $\Omega_{\alpha}$. The direction of ${\bf B}$ is then 
purely rational and all the corresponding open 
orbits (if they exist) should be periodic in $R^{3}$. However,
the mean directions of these open orbits can be different
in this case for the different planes orthogonal to the
magnetic field and the corresponding contributions to
conductivity can not be then written in the form 
(\ref{sigoptr}) in the same coordinate system. The
conductivity tensor $\sigma^{ik}$ can have then the full rank
in the limit $B\tau \rightarrow \infty$ and the conductivity
remains constant for all the directions in $R^{3}$ in the
strong magnetic field limit. This situation, however, is
completely unstable and disappear after any small rotation
of ${\bf B}$. 

 The second possibility in the case of such directions is that 
all the open trajectories
become singular and form the "singular periodic nets" in the
planes orthogonal to ${\bf B}$ (see Fig. 11). The asymptotic
behavior of conductivity is described then by formula
(\ref{sigcltr}) but is also completely unstable and changes
to (\ref{sigoptr}) after any small rotation of ${\bf B}$.

\begin{figure}
  \begin{picture}(0, 180)
    \leavevmode
    \put(0,0){\epsfxsize=1.0\hsize \epsfbox{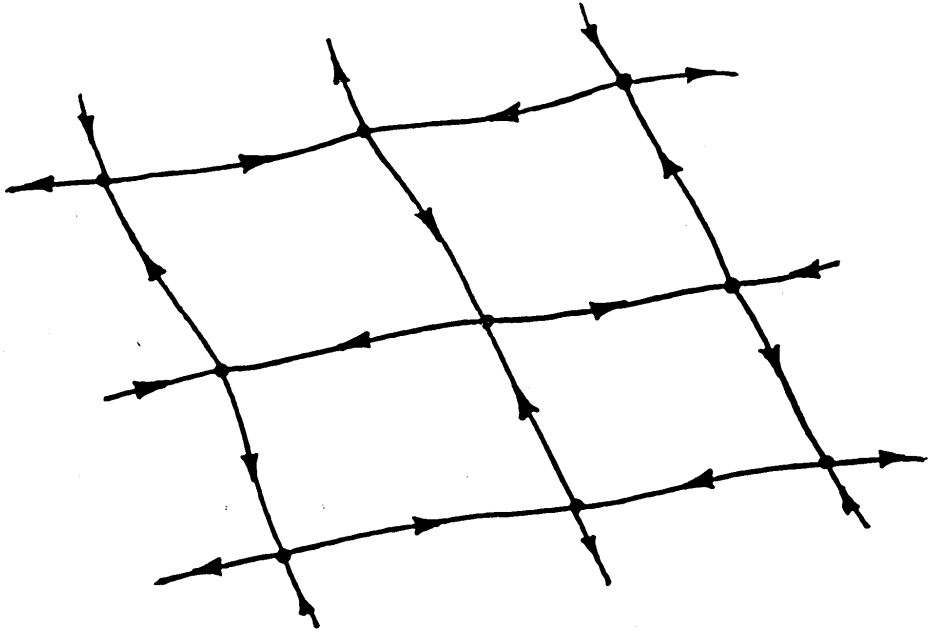}}
  \end{picture}
  \caption{The "singular periodic net" of open trajectories
in the plane orthogonal to ${\bf B}$.}
\end{figure}

 Let us add that the same situations can arise also
on the stable two-dimensional tori $T^{2}_{i}$ in this case
where the directions of open orbits can not be defined anymore
as the intersection of $\Gamma_{\alpha}$ with the plane
$\Pi({\bf B})$. Such, we can have either the "singular nets"
or the regular periodic open orbits on these tori for this
special direction. Also we can have the open
orbits with different mean directions on different tori
in this case but the number of such tori should be $\geq 4$
for any physical type of dispersion relations. This situation
can thus also be observed only for rather complicated Fermi 
surfaces. 

 Let us mention also that for the directions of ${\bf B}$ 
close to this special rational directions the widths of the
straight strips containing the regular open orbit can become
very big in the planes $\Pi({\bf B})$ (see Fig. 12). So from 
physical point of view the conductivity phenomena will not 
"feel" the mean directions of open orbits for the directions 
of ${\bf B}$ close enough to these special points
even for rather big (but finite) values of magnetic field.
Instead, the oscillations of the trajectory within the strip
(Fig. 12) will be essential for conductivity up to the rather
big values of $B$ such that $\omega_{B}\tau \sim L/p_{0}$
(where $L$ is the width of the strip and $p_{0}$ is the size
of the Brilluone zone). These situation exists, however, if
the open orbits with corresponding direction exist also for
${\bf B} = {\bf B}_{0}$ where ${\bf B}_{0}$ is the special
rational direction in the stability zone. For ${\bf B}$
close to ${\bf B}_{0}$ we will observe then exactly this 
direction up to the values of $B$ such that 
$\omega_{B}\tau \sim L/p_{0}$ and then the regime will change 
to the common situation corresponding to given stability zone.
Experimentally we will observe then the "small spots" around 
these directions on the unit sphere where the anisotropy of
$\sigma^{ik}$ corresponds to the direction of orbits for
${\bf B} = {\bf B}_{0}$. In the most complicated case when we 
have the open orbits with different mean directions for
${\bf B} = {\bf B}_{0}$ we will have then the finite conductivity
for all the directions in $R^{3}$ in the corresponding spot
for rather big values of $B$.

 For the "special rational directions" corresponding to the
"singular net" on the stable tori $T^{2}_{i}$ the behavior
of $\sigma^{ik}$ will correspond to the common form for
a given stability zone even for ${\bf B}$ very close to 
${\bf B}_{0}$. However, the measure of open orbits will
tend to zero as ${\bf B} \rightarrow {\bf B}_{0}$ in the
stability zone. The dimensionless coefficients ($*$) in the 
formula (\ref{sigoptr}) will vanish then for 
${\bf B} \rightarrow {\bf B}_{0}$ although the integral plane
$\Gamma_{\alpha}$ will be observable up to 
${\bf B} = {\bf B}_{0}$. 

 We mention here at last that both cases when 
${\bf B}^{\alpha}_{0}$
belongs or does not belong to the corresponding stability zone
$\Omega_{\alpha}$ are possible in the examples.

\begin{figure}
  \begin{picture}(0, 170)
    \leavevmode
    \put(0,0){\epsfxsize=1.0\hsize \epsfbox{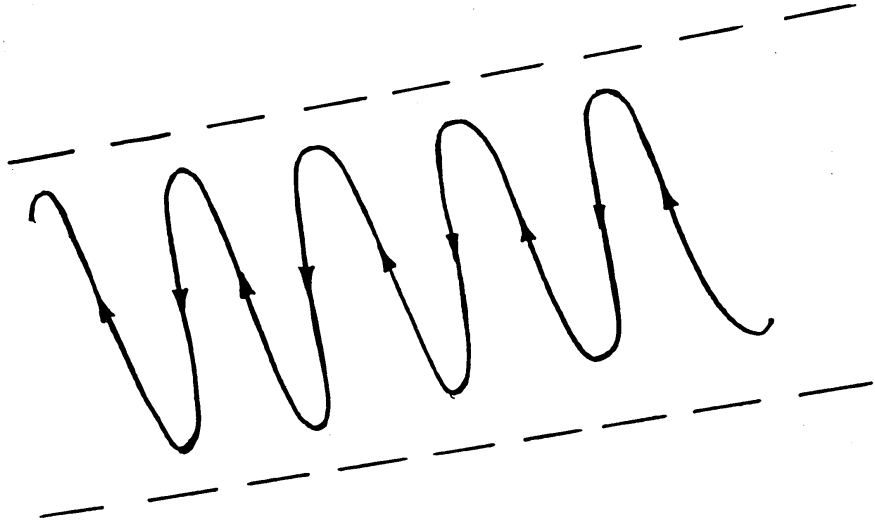}}
  \end{picture}
  \caption{The wide finite strip in the plane $\Pi({\bf B})$
containing the open orbits for ${\bf B}$ close to 
"Special rational direction" within the stability zone.}
\end{figure} 
 
 Let us describe also the situation on the boundary of the 
stability zone on the unit sphere. We will consider the simplest
case of the Fermi surface on Fig. 5,c but the same situation
will take place also for general case. Let us consider the 
sections of this Fermi surface by planes orthogonal to ${\bf B}$
where direction of ${\bf B}$ is close to normal direction of
the warped planes (see Fig. 9, a).

 We have then the two parallel planes (with holes)
in ${\bf p}$-space corresponding to two tori $T^{2}$
in $T^{3}$ divided by cylinder of closed trajectories. The set of
compactified carriers of open orbits ${\bar S}_{F}$ on 
the Fermi level is just the pair
of parallel tori in this case and the height of the 
cylinder of closed trajectories depends on the direction
of magnetic field ${\bf B}$. For the directions of ${\bf B}$
belonging to the boundary of the stability zone this
height is zero and the corresponding cylinder becomes
just the singular orbit connecting two critical points. 
For the directions of ${\bf B}$ 
far enough from the normal
direction the trajectories can "jump" from one plane to
another and the stable open orbits will be destroyed.
This case describes well the general situation on the boundary
of stability zone for generic Fermi surfaces. Namely, we have
generically two tori covered by the open orbits 
divided by cylinder of closed trajectories
near the boundary of the stability zone.
The height of the cylinder becomes equal to zero at the boundary
of the stability zone and then the tori disappear giving the new
reconstruction of Fermi surface ${\bar S}_{F}$. The phase
space occupied by the open orbits thus remains constant  
near the boundary curves and the corresponding 
conductivity tensor (\ref{sigoptr}) has the finite values of
coefficients $*$ up to the boundary of the stability zone.

\vspace{0.5cm}

 Let us discuss now in more details the situation of "chaotic
open" electron orbits which can appear for the cases of
${\bf B}$ with irrationality $3$ (\cite{dynn4}). 
The approximate form of 
trajectories of such kind is represented on Fig. 13.

\begin{figure}
  \begin{picture}(0, 170)
    \leavevmode
    \put(0,0){\epsfxsize=1.0\hsize \epsfbox{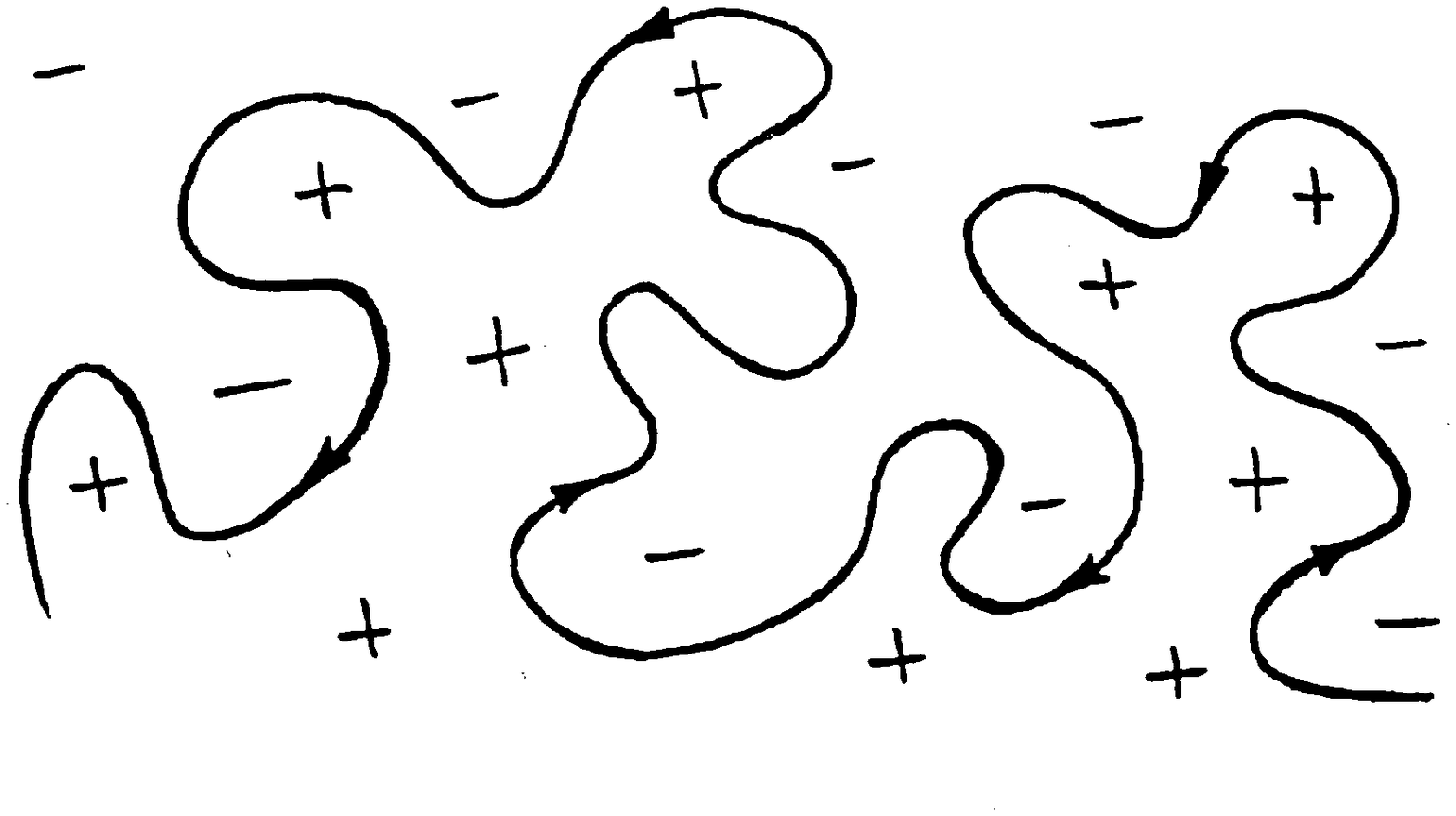}}
  \end{picture}
  \caption{The chaotic open orbit of Dynnikov type. The signs
$"+"$ and $"-"$ denote the regions with $\epsilon > \epsilon_{F}$
and $\epsilon < \epsilon_{F}$ respectively.}
\end{figure}

 For the special examples constructed in \cite{dynn4}
it can be shown (\cite{zorich2}) that there exists a special
direction in $\Pi({\bf B})$ in which the "amplitude of 
oscillations" of trajectories grows faster than in the orthogonal
direction. However, in the general situation this fact is
not strictly proved until now. 

 It can be proved (\cite{dynn7}) that the set of directions
of ${\bf B}$ corresponding to such kind of trajectories 
has the measure zero on $S^{2}$
if we restrict ourselves on the fixed generic Fermi level
$\epsilon({\bf p}) = \epsilon_{F}$. This means actually that
the experimentally observable set of such directions of 
${\bf B}$ has generically measure zero for the case of normal
metals though they still can be found for some special 
directions. 

 Nevertheless, the total set of such directions for the whole 
dispersion relation has a very complicated (fractal type) 
structure and its total measure is still unknown in the general 
situation. 

\vspace{0.5cm}

 {\bf Conjecture.} (S.P.Novikov).

 The total set of the directions of ${\bf B}$ corresponding
to the chaotic behavior has the measure $0$ for the whole 
generic dispersion relation and the Hausdorf dimension strictly 
less than $2$.

\vspace{0.5cm}

 The contribution to the conductivity tensor for the type of 
chaotic trajectories described above is rather different 
from both the cases (\ref{sigcltr}) and (\ref{sigoptr})
(\cite{malts}). Let us mention first of all that the
component of ${\bar S}_{F}$ on the Fermi level carrying the 
trajectory of this kind should have at least genus $3$. 
This means for instance that for any Fermi surface $S_{F}$
with the genus less than $6$ we can have only one such
component (homologous to zero) in the three-dimensional torus 
$T^{3}$. For any physical dispersion relation 
($\epsilon({\bf p}) = \epsilon(-{\bf p}))$ we will have then 
that this component should be invariant w.r.t. to the inversion 
transformation ${\bf p} \rightarrow - {\bf p}$ for the 
appropriate initial point ${\bf p}_{0} = 0$. It follows 
from this symmetry that the average of any component of group
velocity ${\bf v}_{gr}$ over this part of the Fermi surface
should be zero for any physical Fermi surface of genus 
$\leq 6$. The fact that the chaotic trajectory is everywhere
dense on this component leads then to the fact that the 
contribution to the conductivity tensor disappears as
$B\tau \rightarrow \infty$ for such type of trajectories
in all the directions in $R^{3}$ including the direction of
magnetic field ${\bf B}$. The asymptotic form of $\sigma^{ik}$
as $B\tau \rightarrow \infty$ was suggested in ~\cite{malts}
and contains the fractional powers of parameter 
$\omega_{B}\tau$ decreasing as $B \rightarrow \infty$.
Let us just write here the corresponding form in the coordinates
with ${\bf B}$ directed along the $z$ axis and the $x$ and $y$
axes oriented in a special way connected with some aspects
of geometry of the chaotic orbit:

\begin{equation}
\label{sigsttr}
\sigma^{ik} \simeq {n e^{2} \tau \over m^{*}} \,
\left( \begin{array}{ccc}
(\omega_{B}\tau)^{-2\alpha} & (\omega_{B}\tau)^{-1} &
(\omega_{B}\tau)^{-\alpha -\gamma} \cr
(\omega_{B}\tau)^{-1} & (\omega_{B}\tau)^{-2\beta} & 
(\omega_{B}\tau)^{-\beta -\gamma} \cr
(\omega_{B}\tau)^{-\alpha -\gamma} & 
(\omega_{B}\tau)^{-\beta -\gamma} & 
(\omega_{B}\tau)^{-2\gamma} + T^{2}/\epsilon_{F}^{2}
\end{array} \right)
\end{equation}
where $0 < \alpha, \beta, \gamma < 1$, $\alpha + \beta = 1$ 
and $T/\epsilon_{F}$ is a parameter of order of $10^{-4}$.
The contribution (\ref{sigsttr}) should be added with the
contribution of the closed orbits having the form 
(\ref{sigcltr}). The conductivity along the magnetic field
${\bf B}$ will then be finite at 
$\omega_{B}\tau \rightarrow \infty$ due to the presence of
closed trajectories. However, this conductivity should have 
the local minimum on the angle diagram in this case since the
finite part of the Fermi surface will be excluded from it
in the strong magnetic field limit.

 We can describe now the total picture for the angle diagram
of conductivity in normal metal in the case of geometric
strong magnetic field limit. 
Namely, we can observe the following objects on the
unit sphere parameterizing the directions of ${\bf B}$:

\vspace{0.5cm}

 1) The "stability zones" $\Omega_{\alpha}$ corresponding to 
some integral planes $\Gamma_{\alpha}$ in the reciprocal lattice
("Topological Quantum numbers"). This "Topological Type" of open 
trajectories
is stable with respect to small rotations of ${\bf B}$ and this
is the only open orbits regime which can have a non-zero measure
on the unit sphere. All the "stability zones" have the piecewise 
smooth boundaries on $S^{2}$ and are given by the condition

$$\epsilon_{1}({\bf B}) \,\,\, \leq \,\,\, \epsilon_{F} \,\,\,
\leq \,\,\, \epsilon_{2}({\bf B})$$
where $\epsilon_{1}({\bf B})$, $\epsilon_{2}({\bf B})$ are
the piecewise smooth functions defined in the Theorem $3$.
Generally speaking, this set is not everywhere dense
anymore being just a subset of the corresponding
set for the whole dispersion relation and have some
rather complicated geometry on the unit sphere.

 The corresponding behavior of conductivity
is described by the formula (\ref{sigoptr}) and reveals the strong
anisotropy in the planes orthogonal to the magnetic field.
For rather complicated Fermi surfaces we can observe also the
"sub-boundaries" of the stability zones where the coefficients
in (\ref{sigoptr}) have the sharp "jump".

\vspace{0.5cm}

 2) The net of the one-dimensional curves containing directions
of irrationality $\leq 2$ where the additional two-dimensional
tori (homologous to zero in $T^{3}$) can appear. The corresponding
parts of the net are always the parts of the big (passing
through the center of $S^{2}$) circles orthogonal to some
reciprocal lattice vector where the condition

$$\epsilon_{1}^{\prime}({\bf B}) \,\,\, \leq \,\,\, \epsilon_{F} 
\,\,\, \leq \,\,\, \epsilon_{2}^{\prime}({\bf B})$$
is satisfied. The asymptotic behavior of conductivity is given
again by the formula (\ref{sigoptr}).

 Let note also that these special curves on $S^{2}$ can be
considered actually as the reminiscent of the bigger stability
zones if we don't restrict ourselves just by one Fermi surface.
The structure of such sets thus can be used to get more information 
about the corresponding total structure for the whole dispersion 
relation in metal. In particular, the mean direction of such open 
orbits coincides with the mean direction of the generic open
orbits in the intersections of the net with the stability
zones (except the "Special rational directions"). The corresponding
conductivity tensor is given then by the same formula (\ref{sigoptr})
where the dimensionless coefficients "jump" on the curves of the
net.

\vspace{0.5cm}

 3) The "Special rational directions". 

 Let us remind that we call the special 
rational direction the direction of ${\bf B}$
orthogonal to some plane $\Gamma_{\alpha}$ in case when this
direction belongs to the same stability
zone on the unit sphere. We can have here
all the possibilities described earlier for this situation
(i.e. regular behavior with vanishing coefficients in
(\ref{sigoptr}), spots with isotropic or anisotropic behavior
of conductivity different from the given by corresponding 
"Topological quantum numbers", "partly stable" 
isotropic or anisotropic addition to the (\ref{sigoptr}), etc.)

\vspace{0.5cm}

 4) The chaotic open orbits of Tsarev type 
(${\bf B}$ of irrationality $2$).

 We can have points on the unit sphere where the open orbits 
are chaotic in Tsarev sence. All open trajectories still have
the asymptotic direction in this case and the conductivity
reveals the strong anisotropy in the plane orthogonal to 
${\bf B}$ as $B \rightarrow \infty$. The $B$ dependence, however
can be slightly different from the formula (\ref{sigoptr}) in
this case.

\vspace{0.5cm}

 5) The chaotic open orbits of Dynnikov type 
(${\bf B}$ of irrationality $3$). 

 For some points on $S^{2}$ we can have the chaotic open 
orbits of Dynnikov type on the Fermi surface. 
At these points the local
minimum of conductivity along the magnetic field is expected.
The conductivity along ${\bf B}$ however remains finite as
$B \rightarrow \infty$ in general situation because of the 
contribution of closed trajectories. The main asymptotic of
conductivity tensor in plane orthogonal to ${\bf B}$ can
then be extracted from formula (\ref{sigsttr}) in strong
magnetic field limit.

\vspace{0.5cm}

 6) At last we can have the open regions on the unit sphere
where only the closed trajectories on the Fermi level are 
present. The asymptotic behavior of conductivity tensor is 
given then by the formula (\ref{sigcltr}).

\vspace{0.5cm}

 Let now point out some new features connected with the 
"magnetic breakdown" (self-intersecting Fermi surfaces) which
can be observed for rather strong magnetic fields. Up to this
point it has been assumed throughout that different parts
of the Fermi surface do not intersect with each other. 
However, it is possible for some special lattices that the
different components of the Fermi surface (parts corresponding
to different conductivity bands) come very close to each other
and may have an effective "reconstruction" as a result of
the "magnetic breakdown" in strong magnetic field limit.
In this case we can have the situation of the electron
motion on the self-intersecting Fermi surface such that
the intersections with other pieces do not affect at all the 
motion on one component. (The physical conditions for the
corresponding values of $B$ can be found in \cite{etm}).
In this case the picture described above should be considered
independently for all the non-selfintersecting pieces 
of Fermi surface and we can have simultaneously several 
independent angle diagrams of this form on the unit sphere.
Such we can have here the overlapping stability zones
where the open orbits can have different mean directions.
The correspondent conductivity tensor will then be given
just as a sum of all conductivity tensors corresponding
to different non-selfintersecting components. 
(The problem of the magnetic breakdown was brought to
the authors' attention by M.I.Kaganov.)

 Let us mention also that in the paper ~\cite{dynmal} the 
possibilities of the investigation of total topological 
characteristics of whole dispersion relation $\epsilon({\bf p})$
were discussed. In particular the behavior of electrons
injected in the empty band of semiconductor in the presence of
magnetic field was considered. However, the corresponding
magnetic fields should be extremely high $(\sim 10^{2} t)$
in this case which make such experiments very difficult.
Most probably this situation should be considered now
just as theoretical possibility.

 Coming back to the old results we can see now that statement of 
I.M.Lif\-shitz an V.G.Peschanski concerning the "thin spatial net"
(Fig. 3, a,b) can be considered as the particular case of
general topological theorem with the simplest set of planes
$\Gamma_{1}$, $\Gamma_{2}$, $\Gamma_{3}$ given by the coordinate
planes $xy$, $yz$ and $xz$ (i.e. types 
$(\pm 1,0,0), (0,\pm 1,0), (0,0,\pm 1)$ only). 
We predict the existence of the
integral planes also in the case of analytic dispersion
relation (\ref{anferms}) for all the stability zones shown on
Fig. 4. As then follows from the rotational symmetry of
the Fermi surfaces the integral 
planes $\Gamma_{i}$ in this case
should be orthogonal to the unit vectors directed to the centers
of zones shown at Fig. 4 from the center of unit sphere.
This situation, however, is not necessary in the general case
and caused here by the symmetry of the dispersion relation
for the biggest stability zones. All centers of zones at 
Fig. 4 are the "Special rational directions" of the magnetic
field in this case. As again follows from rotational symmetry
the corresponding structure of orbits 
can be only the "singular nets" 
in the planes orthogonal to ${\bf B}_{0}$. The conductivity 
tensor should thus correspond to the form (\ref{sigoptr}) 
with the same "Topological quantum numbers" everywhere and 
the vanishing dimensionless coefficients $*$ near the centers of 
zones. The one-dimensional curves on the Fig. 4 correspond in this
case to the non-generic directions of ${\bf B}$ with the
"partly stable" periodic open trajectories.

 As we have already mentioned we expect here also the infinite 
number of smaller stability zones (as well as the chaotic 
regimes) with the infinite number of corresponding integral 
planes $\Gamma_{\alpha}$. The geometry of the zones and the 
corresponding integer numbers 
$(n^{\alpha}_{1}, n^{\alpha}_{2}, n^{\alpha}_{3})$ will depend
on parameters $\alpha$, $\beta$, $\delta$ and $\zeta_{0}$ in
(\ref{anferms}) being defined by geometry of the correspondent
Fermi surface. The smaller stability zones correspond in the
general picture to bigger numbers 
$(n^{\alpha}_{1}, n^{\alpha}_{2}, n^{\alpha}_{3})$ 
and are more sensitive to the change of parameters in the
analytic approximation.

\vspace{0.5cm}

 The early experimental investigations of different  
conductivity behavior in strong magnetic fields can be found
in \cite{ag1}-\cite{gaid}. 
Let us consider at the end the experimental 
results for conductivity in the plane orthogonal to ${\bf B}$ for
$Au$ obtained by Yu.P.Gaidukov in 1959 (\cite{gaid}).

 For the comparison with the experimental data we give here
the asymptotic form of the resistance tensor, inverse to $\sigma$:\,\,
$R = \sigma^{-1}$ in the same basis as $\sigma$ above 
(see ~\cite{etm},~\cite{abr}).

{\bf Case 1.} (Closed orbits).

\begin{equation}
\label{resis1}
{\hat R} \simeq {m^{*} \over ne^{2}\tau}
\left( \begin{array}{ccc}
1 & \omega_{B}\tau & 1 \cr
\omega_{B}\tau & 1 & 1 \cr
1 & 1 & 1 \end{array}
\right)
\end{equation}
(Part of matrix proportional to $B$ is skew-symmetrical).

{\bf Case 2.} Generic open orbits.

\begin{equation}
\label{resis2}
{\hat R} \simeq {m^{*} \over ne^{2}\tau}
\left( \begin{array}{ccc}
(\omega_{B}\tau)^{2} & \omega_{B}\tau & \omega_{B}\tau \cr
\omega_{B}\tau & 1 & 1 \cr
\omega_{B}\tau & 1 & 1 \end{array}
\right)
\end{equation}

 As we can see from
(\ref{resis2}), we should observe $B^{2}$ - dependence
of the resistance
$\rho \sim (B^{2} cos^{2}\alpha)\rho_{0}$
in the plane orthogonal to ${\bf B}$ in the Case 2.
The coefficient
$(cos^{2}\alpha)$ is equal to $1$ for the electric field directed
along the vector ${\bf e}_{1}$ - zero eigenvector
for conductivity tensor (\ref{sigoptr}) in the plane orthogonal to
${\bf B}$. The same behavior will also be qualitatively true for
the case of Tsarev chaotic open orbits with slightly different
asymptotic in $B$.

 Let us omit here the expression for resistivity tensor 
corresponding to more general Dynnikov chaotic trajectories 
(~\cite{malts}) and write here just the expression for resistivity 
in the plane orthogonal to ${\bf B}$. According to the conjecture
in ~\cite{malts} we should have the "scaling behavior" of
the resistance $\rho \sim B^{\alpha}$ in this situation
where $1 < \alpha < 2$ for the generic direction in $\Pi({\bf B})$.

 On the Fig.14 (Fig.11 in ~\cite{gaid}) we can see series
of black domains where $B^{2}$ - dependence has been observed.

\begin{figure}
  \begin{picture}(0, 280)
    \leavevmode
    \put(0,0){\epsfxsize=1.0\hsize \epsfbox{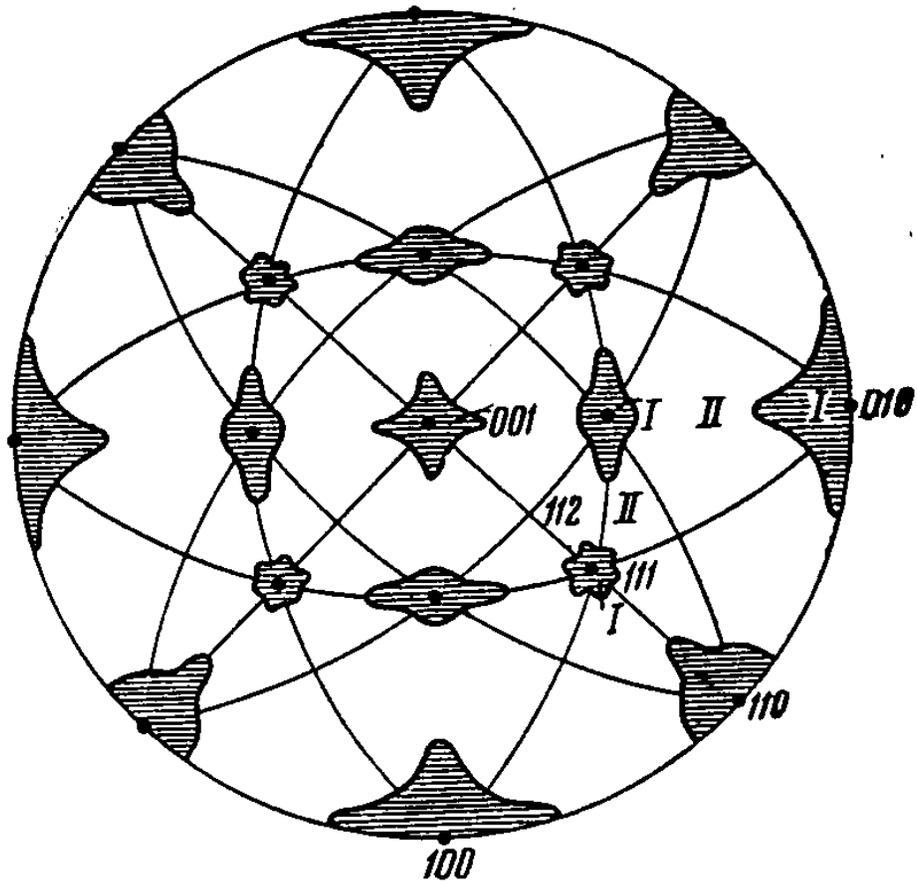}}
  \end{picture}
\caption{Experimental data obtained by Yu.P.Gaidukov for Au. Black
domains correspond to Case 2. }
\end{figure}

 It is interesting to point out that in the centers of the black
domains we find dots where the resistance has "very deep minima"
and correspond to the Case 1, according to the results of 
~\cite{gaid}. This situation is precisely analogous to the
case of the biggest stability zones for the analytic dispersion
relation (\ref{anferms}) and we have the "singular nets"
in the planes orthogonal to these "Special rational directions".

 The resistance within the black domains should be the
$B^{2}$ - type, like in the Case 2. However, this dependence
was found experimentally like $B^{\alpha}$ for $\alpha \leq 2$
("slightly less", as written in ~\cite{gaid}). Probably,
magnetic field $B \sim 1 T$ was not enough for our asymptotic
behavior. We think that
these experiments should be repeated with $B \sim 10 T$
to get the precise $B^{2}$-dependence.
In this case we can definitely state that
these black domains are really the "Stability Zones"
whose topological types correspond to the integral planes   
orthogonal to unit vectors directed into the central points.
So we predict here the following set of the 
"Topological Quantum numbers": 

$$(M_{1}, M_{2}, M_{3}) = (\pm 1, 0, 0), (0, \pm 1, 0),
(0, 0, \pm 1), (\pm 1, \pm 1, 0),$$
$$(\pm 1, 0, \pm 1), (0, \pm 1, \pm 1), (\pm 1, \pm 1, \pm 1)$$

 Everywhere in the white area we have the Case 1 and the
finite resistance as $B \rightarrow \infty$ in $\Pi({\bf B})$. 

 The interesting behavior was found also along 
the black lines on the Fig 14.

 Namely, in many points of these lines the resistance has very
deep maxima where the asymptotic behavior like $B^{\alpha}$ has 
been observed for the different values of $\alpha$: 
$1 < \alpha < 1.8$.
In our opinion, these experimental results should be improved for
$B \geq 10 T$ instead of $B \sim 2 T$ like in ~\cite{gaid}.

 {\bf Conjecture.}
The non-generic chaotic orbits exist here.

 Unfortunately the conductivity along the magnetic field ${\bf B}$
was not measured in ~\cite{gaid} so we do not have any information
if any local minima of longitudinal conductivity were observed for
these special directions.

\end{document}